\documentclass[useAMS,usenatbib]{mn2e}

\include{epsf}

\usepackage{graphicx}

\title[XMM/SDSS: The clustering of X-ray AGN at z=0.1]{The clustering of X-ray selected AGN at z=0.1}
\author[G. Mountrichas \& A. Georgakakis]{G. Mountrichas, A. Georgakakis
\\ \\  
  National Observatory of Athens,  V.  Paulou \&  I.  Metaxa,
  11532, Greece\\   
}

\begin{document}



\maketitle

\label{firstpage}

\begin{abstract}
The clustering properties of moderate luminosity ($L_X = \rm 10^{41} -
10^{44}  \, erg  \,s^{-1}$)  X-ray selected  AGN  at $z\approx0.1$  are
explored.   X-ray sources  in the  redshift interval  $0.03<z<0.2$ are
selected  from  a  serendipitous  XMM  survey of  the  SDSS  footprint
(XMM/SDSS) and are cross-correlated  with the SDSS Main galaxy sample.
The  inferred X-ray AGN  auto-correlation function  is described  by a
power   law  with   amplitude  $r_0\approx5\,$h$^{-1}$Mpc   and  slope
$\gamma\approx2.0$.  The corresponding mass  of the dark matter haloes
that host X-ray AGN at  $z\approx0.1$ is $\approx 10^{13} \,h ^{-1} \,
M_{\sun}$.  Comparison with studies at higher redshift shows that this
mass scale is  characteristic of moderate luminosity X-ray  AGN out to
$z\approx  1$.  Splitting the  AGN sample  by rest-frame  colour shows
that  X-ray  sources  in  red  hosts are  more  clustered  than  those
associated  with   blue  galaxies,   in  agreement  with   results  at
$z\approx1$. We also find that the host galaxies of X-ray AGN have
lower stellar masses compared to the typical central galaxy  of a
$\approx  10^{13} \,h  ^{-1} \,  M_{\sun}$ dark matter halo. AGN hosts
either have experienced less stellar mass growth compared to the average
central galaxy of a $\approx  10^{13} \,h  ^{-1} \,  M_{\sun}$ halo or a
fraction of them are associated with satellite galaxies. 
\end{abstract}

\begin{keywords}
galaxies: active, galaxies: haloes, galaxies: Seyfert, X-rays: diffuse
background 
\end{keywords}

\section{Introduction}

Observations (e.g.  Ferrarese \& Merritt  2000, Gebhardt et  al. 2000)
and theoretical arguments (e.g.  Silk 1998, King 2005) suggest that the
growth of supermassive black holes  (SMBHs) at the centres of galaxies
are intimately related  to the formation of their  hosts.  However, the
details of  the interplay between  these two components are  still not
well understood.

The environment of AGN, i.e. the  mass of the typical dark matter halo
(DMH) in which they live, is a powerful diagnostic of the physics that
drive the  formation of SMBHs  and their hosts.  Different  models for
the  co-evolution   of  AGN   and  galaxies  make   specific  testable
predictions  on  how  the  environment  of active SMBHs  depends  on
redshift and accretion luminosity (e.g.  Marulli et al.  2006, Hopkins
et  al.  2007,  Fanidakis 2010,  Degraf et  al.  2010,  Bonoli  et al.
2009).  Moreover, there has been progress recently in phenomenological
methods to associate baryons to  DMHs at different redshifts, to infer
in  a least  model  dependent  way how  galaxies  evolve at  different
environments (e.g. Zheng  et al. 2007, Conroy \&  Wechsler 2009, Zehavi
et  al.   2011, Avila-Reese  \&  Firmani  2011).   By determining  the
typical DMH of AGN one can  place them in the context of those studies
to explore the  conditions under which SMBH grow  at different epochs.
An advantage  of studying the environment  of AGN is  that it provides
one of the few diagnostics  of the AGN/galaxy interplay that is immune
to contamination of the stellar light of galaxies by emission from the
central engine. AGN can easily outshine their hosts, thereby rendering
studies of the stellar  mass, star-formation history and morphology of
their  hosts  challenging and  prone  to  systematics.   This poses  a
serious limitation  in the  comparison between observations  and models
for the  growth of SMBHs  and highlights the importance  of clustering
studies.

Powerful UV bright QSOs are one of the few AGN classes for which tight
constraints on  their large scale distribution are  available. This is
because of  the apparent  brightness of these  sources in  the optical
which has allowed large  spectroscopic follow-up programs, such as the
2QZ (2dF QSO Redshift Survey, Croom et al.  2004), SDSS (Sloan Digital
Sky Survey, Richards  et al.  2002, Schneider et  al.  2005) and 2SLAQ
(2dF-SDSS  LRG and  QSO  Survey, Cannon  et  al.  2006,  Croom et  al.
2008).  The  picture emerging from  those studies is that  powerful UV
bright QSOs live in dark matter haloes of few times $10^{12}\,h^{-1}\,
M_{\odot}$  almost independent  of redshift  and  accretion luminosity
(e.g.  Croom  et al.  2005, Myers  et al.  2007,  da $\hat{A}$ngela et
al.   2008, Ross  et  al.  2009,  Ivashchenko  et al.   2010, but  see
Mountichas  et al.   2009).  These  properties are  broadly consistent
with the predictions of the merger  model for the growth of SMBH (e.g.
Hopkins et al.   2007, Bonoli et al.  2009),  suggesting that powerful
QSOs are the product of interactions between gas rich galaxies.

The QSOs selected in the above surveys however, represent luminous and
rare sources above the knee of the AGN luminosity function. Therefore,
they are  not representative of  the overall AGN population  and their
contribution  to  the  accretion  power  of  the  Universe  is  small.
Additionally,  their   selection  at  UV/optical   wavelengths  raises
concerns on possible biases  against obscured sources. Observations at
X-ray wavelengths  provide an  efficient way of  selecting AGN  over a
wide luminosity baseline nearly  independent of obscuration.  Study of
the  clustering  of X-ray  AGN  can  therefore  constrain the  fueling
mechanism of  the sources that  dominate the accretion history  of the
Universe. Despite considerable progress in the last few years however,
the  large  scale  distribution  of   X-ray  AGN  is  still  not  well
constrained.  This  is primarily because of the  apparent faintness of
these sources at optical wavelengths and their relatively high surface
density on the sky, which make large spectroscopic follow-up programs,
similar to those  carried out for UV bright  QSOs, extremely expensive
in telescope time.

In the  absence of  any follow-up observations,  the most  widely used
approach  for studying  the clustering  of  X-ray AGN  is the  angular
auto-correlation function (e.g. Basilakos  et al.  2004, 2005, Plionis
et al. 2008, Ebrero et al. 2009). The main limitation of this approach
is that  assumptions have to be  made on the  redshift distribution of
the AGN population to infer the mass of their dark matter haloes. This
introduces systematics and model  dependent biases which can be large.
This problem  has been mitigated by  extensive follow-up spectroscopic
programs  in selected  few X-ray  survey fields  (e.g.  Barger  et al.
2003, Brusa et al.  2009),  which allowed estimation of the real-space
auto-correlation  function  of  X-ray  AGN samples  (e.g.   Mullis  et
al. 2004,  Gilli et  al. 2005,  2009, Yang et  al.  2006,  Allevato et
al. 2011). One of the results  from these studies is the importance of
sample variance.  Spikes in the redshift distribution of X-ray sources
can significantly affect clustering  studies in current X-ray surveys,
which typically  have small angular  sizes (e.g.  Gilli et  al.  2005,
2009).   An   alternative  approach  for  studying   the  large  scale
distribution of AGN is  the real-space cross-correlation function with
galaxies over the  same cosmological volume (e.g.  Coil  et al.  2009,
Krumpe et  al.  2010).  This  approach has certain merits  compared to
auto-correlation function  methods, as long  as the clustering  of the
galaxy  population  is   known  to  a  good  level   of  accuracy  and
spectroscopy is available for the  AGN.  Firstly, the space density of
galaxies  is   typically  much  larger  than  that   of  AGN,  thereby
suppressing random  errors when counting  AGN/galaxy pairs.  Secondly,
by cross-correlating AGN with galaxies in the same field the impact of
sample variance is minimised. The studies above suggest that X-ray AGN
live in dark  matter haloes with masses $\rm  5\times 10^{12}- 5\times
10^{13}\,h^{-1}\,M_{\odot}$, which are,  on average, more massive than
those of  UV bright QSOs. The  wide range in the  estimated DMH masses
could  be the  result of  random  errors and  systematics that  affect
individual measurements  or because the  AGN clustering may  depend on
the accretion luminosity, the redshift and/or the level of obscuration
of the central engine (e.g. Plionis 2008, Hickox et al. 2011, Allevato
et al. 2011).  

In this  paper we explore changes  in the large  scale distribution of
moderate luminosity ($L_X \rm <10^{44}\, erg \,s^{-1}$) X-ray selected
AGN  from $z\approx1$ to  $z\approx0.1$.  We  use a  serendipitous XMM
survey  of the  SDSS area  (XMM/SDDS, Georgakakis  \& Nandra  2011) to
compile a  sample of low  redshift AGN, $z\approx0.1$.   These sources
are then cross-correlated with the SDSS Main Galaxy sample (Strauss et
al. 2002) to  infer their clustering properties. The  advantage of the
serendipitous XMM/SDSS  survey is that  the AGN selection  function at
low redshift is  almost identical to that of  X-ray AGN at $z\approx1$
detected  in  deep Chandra  and  XMM  surveys. Therefore  differential
selection  effects   between  low   and  high  redshift   samples  are
minimal. This allows direct  comparison of the environment of moderate
luminosity   X-ray  AGN  across   redshift  to   investigate  possible
evolutionary trends.  Throughout this paper  we adopt $H_{0} = \rm 100
\,  km  \,  s^{-1} \,  Mpc^{-1}$,  $\rm  \Omega_{M}  = 0.3$  and  $\rm
\Omega_{\Lambda} =  0.7$.  Rest frame  quantities (e.g.  luminosities,
dark matter halo masses) are parametrised by $h=H_{0} / 100$.

\section{The X-ray AGN sample}

The  clustering   properties  of   X-ray  AGN  at   $z\approx0.1$  are
investigated  by   selecting  low  redshift  sources   detected  in  a
serendipitous  XMM  survey  of  the SDSS  footprint  (XMM/SDSS).   The
construction of the XMM/SDSS  source catalogue, including X-ray source
detection, flux estimation and optical identification, is described in
Georgakakis \& Nandra (2011).

The XMM/SDSS survey includes pointings targeting clusters of galaxies.
The  overdensity of  sources  in  those fields  may  bias large  scale
structure studies. Therefore XMM/SDSS survey fields that have clusters
as  their  prime  targets  are  excluded  from  the  analysis.   These
observations are identified from the  target name keyword of the event
files.   This reduces  the  total XMM/SDSS  survey  area to  102\,$\rm
deg^2$. 

The low  redshift X-ray subsample  of the XMM/SDSS survey  consists of
175 serendipitous hard-band  (2-8\,keV) and 297 full-band (0.5-8\,keV)
detections    with   $0.03<z<0.2$,    X-ray   luminosity    $L_X   \rm
(2-10\,keV)>10^{41} \,  erg \, s^{-1}$ (see  below) and $r<17.77$\,mag
after correcting for Galactic  extinction (Schlegel et al. 1998).  The
magnitude cut corresponds to the  limit of the SDSS Main Galaxy Sample
(Strauss et al. 2002), which provides the majority of redshifts in the
SDSS.   The photometry  is from  the New  York  University Value-Added
Galaxy Catalog  (NYU-VAGC, Blanton et al.  2005)  which corresponds to
the SDSS DR7  (Abazajianet al.  2009). The X-ray  luminosity cut is to
limit the  sample to AGN  which contribute substantially to  the X-ray
luminosity density of  the Universe at low redshift  (e.g. Aird et al.
2010).   As   discussed  by  Georgakakis  et  al.   (2011)
contamination  by normal galaxies  is not  a concern  for luminosities
$L_X(\rm 2-10\, keV) > 10^{41} \, erg \, s^{-1}$.

Clustering results are  presented for both the full  and the hard-band
selected AGN  samples. The  former has the  advantage of  larger size,
thereby  improving the  statistical reliability  of the  results.  The
latter  is  selected  at  rest-frame  energies of  about  2-9\,keV  at
$z=0.1$,  which are  similar to  those of  Chandra and  XMM  X-ray AGN
samples  at $z\approx1$ (typically  1-14\,keV).  This  facilitates the
comparison  across  redshift   by  minimising  differential  selection
effects.

X-ray  luminosities   are  estimated  in  the   2-10\,keV  band  after
correcting the observed flux  in the 2-8\,keV band (hard-band selected
sample)  or  the  0.5-8\,keV  band  (full-band  selected  sample)  for
intrinsic  absorption  parametrised by  the  hydrogen column  density,
$N_H$.  For individual X-ray AGN  this quantity is determined from the
hardness ratios between the  soft (0.5-2\,keV) and the hard (2-8\,keV)
X-ray bands assuming an  intrinsic power-law X-ray spectrum with index
$\Gamma=1.9$ (Nandra et al. 1994).

\section{Methodology}
\subsection{AGN clustering estimation}

Because the  low redshift subset ($0.03<z<0.2$  and $r<17.77$\,mag) of
the XMM/SDSS survey  source catalogue is small, we  choose to quantify
their  clustering  properties  by estimating  their  cross-correlation
function  with  the  much  larger  sample  of  SDSS  spectroscopically
identified galaxies.  We use the NYU-VAGC to select a total of 592,017
sources  in  the  SDSS   Main  Galaxy  spectroscopic  sample  (bitmask
parameter {\sc  vag\_select} equals 7) with redshifts  in the interval
$0.03 - 0.2$ and extinction corrected magnitudes $r<17.77$\,mag.

Incompleteness  in  the galaxy  redshift  catalogue  because of  fiber
collisions is a source of  bias in clustering studies. The SDSS fibers
have finite  size and cannot  be placed closer than  55\,arcsec, which
corresponds  to about  $70\,  h^{-1}  \rm \,  kpc$  at $z=0.1$.   This
separation is too  small to affect our results  and conclusions as the
X-ray/galaxy cross-correlation  signal is  dominated by pairs  on much
larger scales (see Results section).  We nevertheless correct for this
effect by  assigning a source that  was not observed  because of fiber
collisions  the redshift  of the  galaxy  with which  it collided,  as
proposed by Blanton et al. (2005).

The  determination  of the  cross-correlation  function  with a  large
sample   of  galaxies   is   superior  to   the   estimation  of   the
auto-correlation  function in the  case of  small samples,  like X-ray
AGN,   because  random  errors   are  significantly   suppressed.   An
additional advantage is  that the cross-correlation requires knowledge
of the selection function of galaxies only, which is less complex than
that of X-ray AGN. Also, sample  variance is affecting in the same way
X-ray AGN and galaxies. The impact of this bias is therefore minimised
in the cross$-$correlation function calculation. The cross-correlation
approach   however,    requires   an   accurate    estimate   of   the
auto-correlation  function   of  galaxies  to   infer  the  clustering
properties of AGN.

Next we present the equations used to determine the clustering of AGN.
Since  both the auto-correlation  and the  cross-correlation functions
are  special cases of  the 2-point  statistics of  the AGN  and galaxy
populations, they  are both defined  by the same basic  equations.  In
this  section  the term  correlation  function  refers  to either  the
auto-correlation or  the cross-correlation functions.   When necessary
we will differentiate between the two quantities.

The real space correlation function, $\xi$, is calculated as

\begin{equation}\label{eqn:xi}
\xi =\frac{N_{rd}}{N_{gal}}\frac{DD}{DR}-1,
\end{equation}

\noindent where $N_{rd}$ is the  number of random points, $N_{gal}$ is
the number  of galaxies,  $DD$ are the  data-data pairs  at separation
$r$,  i.e.   AGN-galaxy pairs  in  the  cross-correlation function  or
galaxy-galaxy pairs in the case of the auto-correlation function, $DR$
are  the AGN-random pairs  (cross-correlation) or  galaxy-random pairs
(galaxy auto-correlation  function) at separation  $r$.  Random points
within the surveyed area are produced by randominsing the positions of
SDSS galaxies  taking into  account the SDSS  window function  and the
spectroscopic   completeness  of  the   galaxy  sample.    The  random
catalogues provided as  part of the NYU-VAGC data  release version 7.2
are used.  They contain  $\approx 3\times$ more randoms than galaxies.
The random points are distributed with constant surface density in the
SDSS  footprint  as defined  by  the  Large  Scale Structure  mask  of
NYU-VAGC (Blanton et al. 2005).  The redshifts assigned to the randoms
points follow the redshift distribution of the galaxy sample.

When  the  correlation function  is  measured  in redshift-space,  the
clustering is affected at small  scales by the rms velocity dispersion
of AGN along the line of sight and by
dynamical infall of matter into  higher density regions.  If $s_1$ and
$s_2$  are   the  distances   of  two  objects   1,  2,   measured  in
redshift-space, and $\theta$ the angular separation between them, then
$\sigma$ and $\pi$ are defined as

\begin{equation}
\pi=(s_2-s_1), $ along the line-of-sight$,
\end{equation}

\begin{equation}
\sigma=\frac{(s_2+s_1)}{2}\theta , $ across the line-of-sight$.
\end{equation}

\noindent These are small angle approximations. Equation $\ref{eqn:xi}$
then becomes:

 \begin{equation}
\xi(\sigma,\pi) =\frac{N_{rd}}{N_{gal}}\frac{DD(\sigma,\pi)}{DR(\sigma,\pi)}-1,
\label{eqn:wtheta}
\end{equation}

\noindent To   the  first   order,   the  non-linear   redshift-space
distortions, i.e. the small-scale  peculiar velocities, appear only in
the  radial  component.   These  effects are  therefore  minimised  by
integrating $\xi$  along the $\pi$ direction.  The resulting two-point
statistic is the projected correlation function

\begin{equation}
w_p(\sigma)=2\int_0^\infty \xi(\sigma,\pi)d\pi.
\label{eqn:wp}
\end{equation}

\noindent In  practice the maximum scale  of the integration  is set to
$\pi_{max}= \rm 70 \, h ^{-1} Mpc$ (da $\hat{A}$ngela et al. 2008). If
larger scales are included, the  signal will be dominated by noise. If
the  integration is  limited on  small  scales the  amplitude will  be
underestimated.

In  the  estimation  of  the  projected  AGN/galaxy  cross-correlation
function each AGN  is weighted by the factor  $1/V_{max}$, the inverse
of  the maximum volume  within which  a source  can be  detected. This
accounts for the complex selection function of 
the  XMM/SDSS AGN.  Because  of vignetting  and the  spatially varying
width of the XMM's Point  Spread Function (PSF) the solid angle within
which the surveyed area is sensitive to X-ray faint sources is smaller
than  for  brighter  ones.  As  a  result  lower  luminosity  AGN  are
underepresented in the low redshift subset of the XMM/SDSS survey, but
in a  way that can be  accurately quantified and corrected  for in the
$V_{max}$ calculation.  It is emphasised  that this is a  second order
correction  and does  not  have a  strong  impact on  the results  and
conclusions.  For  simplicity  this  weight  is not  included  in  the
equations presented in this section.

Under the assumption that  the real-space correlation function follows
a      power-law      of       the      form      $\xi      (r)      =
\left(\frac{r}{r_o}\right)^{-\gamma}$,   the   real-space  correlation
length, $r_0$, and slope, $\gamma$, can be estimated directly from the
projected correlation function

\begin{equation}
\frac{w_p(\sigma      )}{\sigma}=\left(\frac{r_0}{\sigma}\right)^\gamma
\frac{\Gamma                                        (\frac{1}{2})\Gamma
  (\frac{\gamma-1}{2})}{\Gamma(\frac{\gamma}{2})},
\label{eqn:projected_ro}
\end{equation}

\noindent where $\Gamma  (x)$ is the Gamma function.   This is because
equation \ref{eqn:wp} can be rewritten as

\begin{equation}
w_p(\sigma)=2\int_\sigma^{\pi_{max}}\frac{r\xi    (r)}{\sqrt{r^2-\sigma
    ^2}}dr,
\end{equation}

\noindent and  then solved analytically by  substituting the power-law
form for $\xi(r)$.

The projected  correlation function can  also be used to  estimate the
real-space correlation function,  $\xi (r)$, even if a  power law form
is not adopted. By inverting $w_p(\sigma)$ (Saunders et al. 1992)

\begin{equation}
\xi(r)=-\frac{1}{\pi}\int_r^\infty\frac{d\omega
  (\sigma)/d\sigma}{\sqrt{(\sigma ^2-r^2)}}d\sigma
\label{eqn:xir_1}
\end{equation}

\noindent  and assuming a  step function  for $w_p(\sigma)=w_i$  it is
found

\begin{equation}
\xi(\sigma  _i)=-\frac{1}{\pi}\sum_{j\geq i}\frac{\omega _{j+1}-\omega
  _j}{\sigma  _{j+1}- \sigma  _j}ln{(\frac{\sigma _{j+1}+\sqrt{\sigma_
      {j+1}^2-\sigma    _i^2}}{\sigma    _j+\sqrt{\sigma_   j^2-\sigma
      _i^2}})}
 \label{eqn:xir_2}
\end{equation}

\noindent for $r=\sigma _i$.  That measurement is generally more noisy
than $w_p(\sigma)$.   The projected correlation  function and equation
$\ref{eqn:projected_ro}$  are  therefore   used  to  infer  $r_0$  and
$\gamma$, under  the assumption of  power-law for $\xi$.   However, we
check  that  the  $r_0$  and  $\gamma$  values  obtained  by  equation
$\ref{eqn:projected_ro}$   provide  a   good   approximation  to   the
real-space correlation function infered from equation \ref{eqn:xir_2}.

The  above procedure  allows us  to measure  the  AGN-galaxy projected
cross-correlation  function,  $w_p(AG)$,   and  the  galaxy  projected
auto-correlation  function, $w_p(GG)$.  Using  these measurements  and
assuming a  linear bias, the AGN  projected auto-correlation function,
$w_p(AA)$, is

\begin{equation}
w_p(AA)=\frac{w_p(AG)^2}{w_p(GG)}.
\label{eqn:inferred}
\end{equation}

\noindent  For the estimation  of errors  of the  correlation function
measurements the  survey area  is split into  six subregions,  each of
which   includes   nearly   equal    number   of   X-ray   AGN.    The
cross-correlation  function is  then  estimated for  each  of the  six
subregions. The variance at a given scale is

\begin{equation}
\sigma ^2 = \frac{1}{N-1}\sum_{L=1}^{N}\frac{DR_L}{DR}[\xi_L-\xi]^2,
\end {equation}

\noindent where  N is the  number of fields,  i.e. N=6, $DR_L$  is the
AGN-random  pairs  in  the  field,  $DR$  is  the  overall  number  of
AGN-random pairs,  $\xi_L$ is the  correlation function measured  in a
subregion and $\xi$ is the overall correlation function.

\subsection{Estimation of the bias parameter}

The bias parameter of AGN  relative to the underlying dark matter halo
distribution  is estimated  using  two different  approaches that  are
often  adopted  in  the  literature.   This is  to  facilitate  direct
comparison of our results with previous studies.

The   first   method   uses   the  integrated   projected   AGN/galaxy
cross-correlation  function and  normalizes the  result to  the volume
contained in a sphere with  radius of $20h^{-1} \rm \,Mpc$ (e.g.  Ross
et al.  2009, Mountrichas et al. 2009, da $\hat{A}$ngela et al. 2008)

\begin{equation}
\xi_{20}=\frac{3}{20^3}\int_{r_{min}}^{r_{max}}\xi (r)r^2dr.
\label{eqn:xi20}
\end{equation}

\noindent  where   $\xi_{20}$  is  the   integrated  cross-correlation
function. The lower  and upper limits are set to 1  and $20 h^{-1} \rm
\,  Mpc$,  respectively.   For   the  estimation  of  $\xi_{20}$,  our
$w_p(\sigma)$  measurements  are used.   The  galaxy  bias, $b_G$,  is
estimated as

\begin{equation}
b_G^2=\frac{\xi_{GG}}{\xi_{mm}}\Rightarrow b_G\approx \sqrt\frac{\xi_{20}^{GG}}{\xi_{20}^{mm}},
\end{equation}

\noindent where $\xi_{20}^{mm}$ is the integrated correlation function
of dark matter. It is estimated from the normalized $\Lambda$CDM power
spectrum  model of  Smith et  al. (2003)  for  cosmological parameters
$\Omega _m(z=0)=0.3$,  $\Omega _{\Lambda}(z=0)=0.7$, $\Gamma=0.17$ and
$\sigma_8=0.8$, in accordance with the recent WMAP results (Spergel et
al.  2007). The value  of $\sigma_8$  has been  revised by  the latest
analysis  of  the  WMAP data  from  0.84  to  0.8. To  facilitate  the
comparison  with   previous  studies  we  also   present  results  for
$\sigma_8=0.84$.

Having measured  the galaxy bias we  can then estimate  the AGN bias,
$b_A$, from the relation

\begin{equation}
b_Ab_G=\frac{\xi_{AG}}{\xi_{mm}}\Rightarrow b_A\approx \frac{1}{b_G} \frac{\xi_{20}^{AG}}{\xi_{20}^{mm}},
\end{equation}

\noindent   where  $\xi_{20}^{AG}$   is   the  integrated   AGN/galaxy
cross-correlation  function.  The equation  above  also assumes  scale
independent bias.

The second  approach for estimating  the bias uses the  best-fit $r_0$
and  $\gamma$ values  of the  inferred AGN  auto-correlation function,
obtained   from   equation   $\ref{eqn:inferred}$  (e.g.   Krumpe   et
al. 2010).  The clustering strength is  expressed in terms  of the rms
fluctuation of the density distribution  over a sphere with a comoving
radius of 8h$^{-1}$Mpc

\begin{equation}
\sigma_{8,AGN}^2=J_2(\gamma)\left( \frac{r_0}{8h^{-1}Mpc} \right)^\gamma,
\end{equation}
where 
\begin{equation}
J_2(\gamma)=\frac{72}{(3-\gamma)(4-\gamma)(6-\gamma)2^\gamma}, 
\end{equation}
the AGN bias can be calculated via 

\begin{equation}
b_{AGN}=\frac{\sigma_{8,AGN}}{\sigma_8(z)}.
\label{eqn:bs8}
\end{equation}

\noindent The  bias of AGN or galaxies  is related to the  mass of the
dark matter  halos they live in  (e.g.  Mo \&  White 1996). Therefore
their clustering  properties can  be used to  infer their  dark matter
halo  masses. In this  calculation the  ellipsoidal collapse  model of
Sheth, Mo $\&$ Tormen (2001)  is adotped. The methodology described by
da $\hat{A}$ngela et  al. (2008) and van den  Bosch (2002) is followed
to convert  the bias measurements to  dark matter halo  masses. We use
the bias values estimated from the $\sigma_{8,AGN}$ rms fluctuation of
the density distribution.

\begin{figure*}
\begin{center}
\includegraphics[scale=0.57]{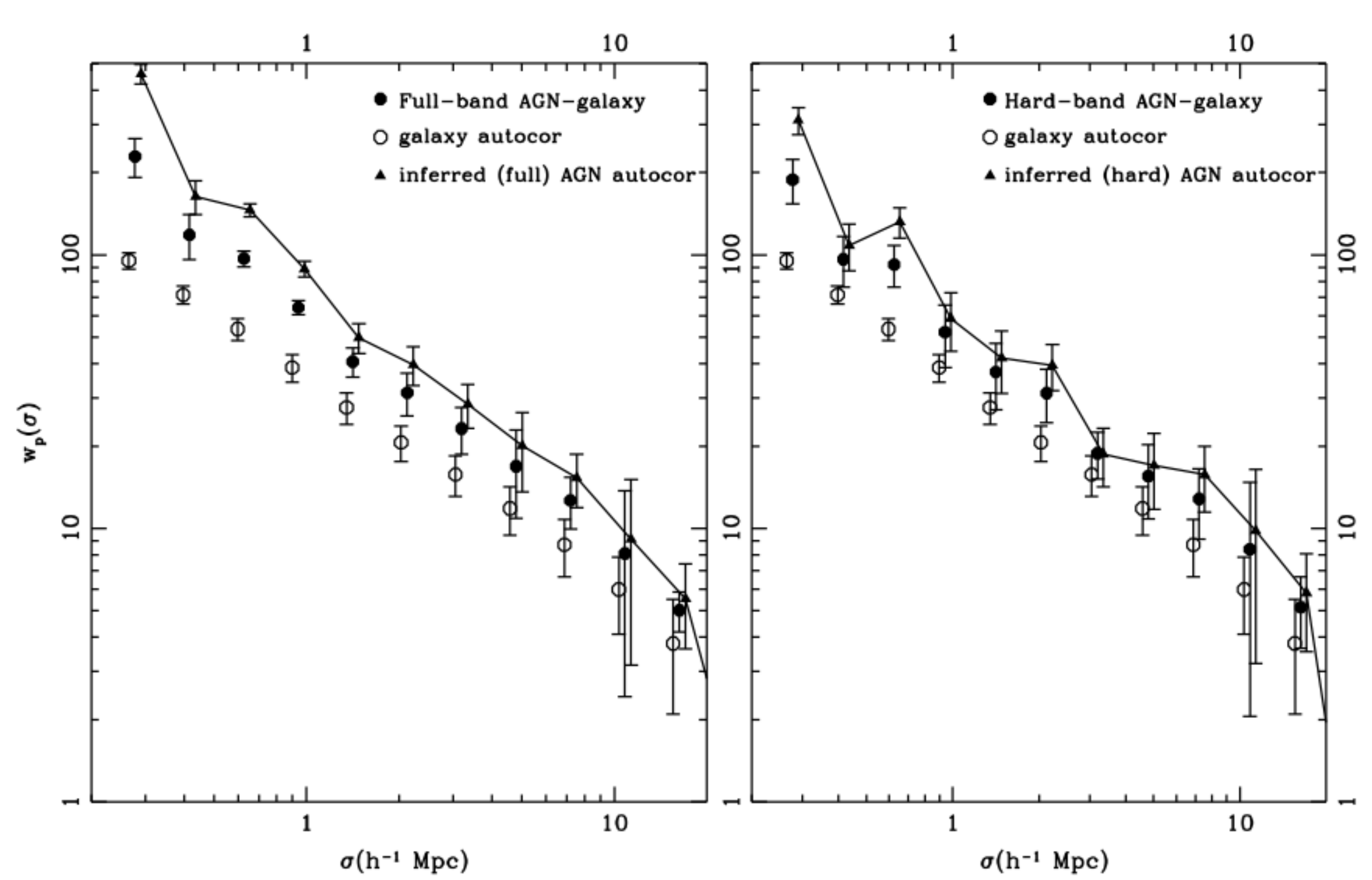}
\caption{Left panel: The full-band AGN-galaxy cross-correlation is shown by filled
  circles.   Open    circles   show   the    galaxy   auto-correlation
  results.  Solid triangles  connected with  the solid  line  show the
  inferred full-band AGN auto-correlation function. Right panel: Same as in the left panel  but using the hard-band AGN sample. For clarity, open circles and solid triangles are offset in the horizontal direction by $\delta \log \sigma = -0.02$ and +0.02, respectively.}
\label{fig:fullband_agn}
\end{center}
\end{figure*}

\begin{figure}
\includegraphics[scale=0.45]{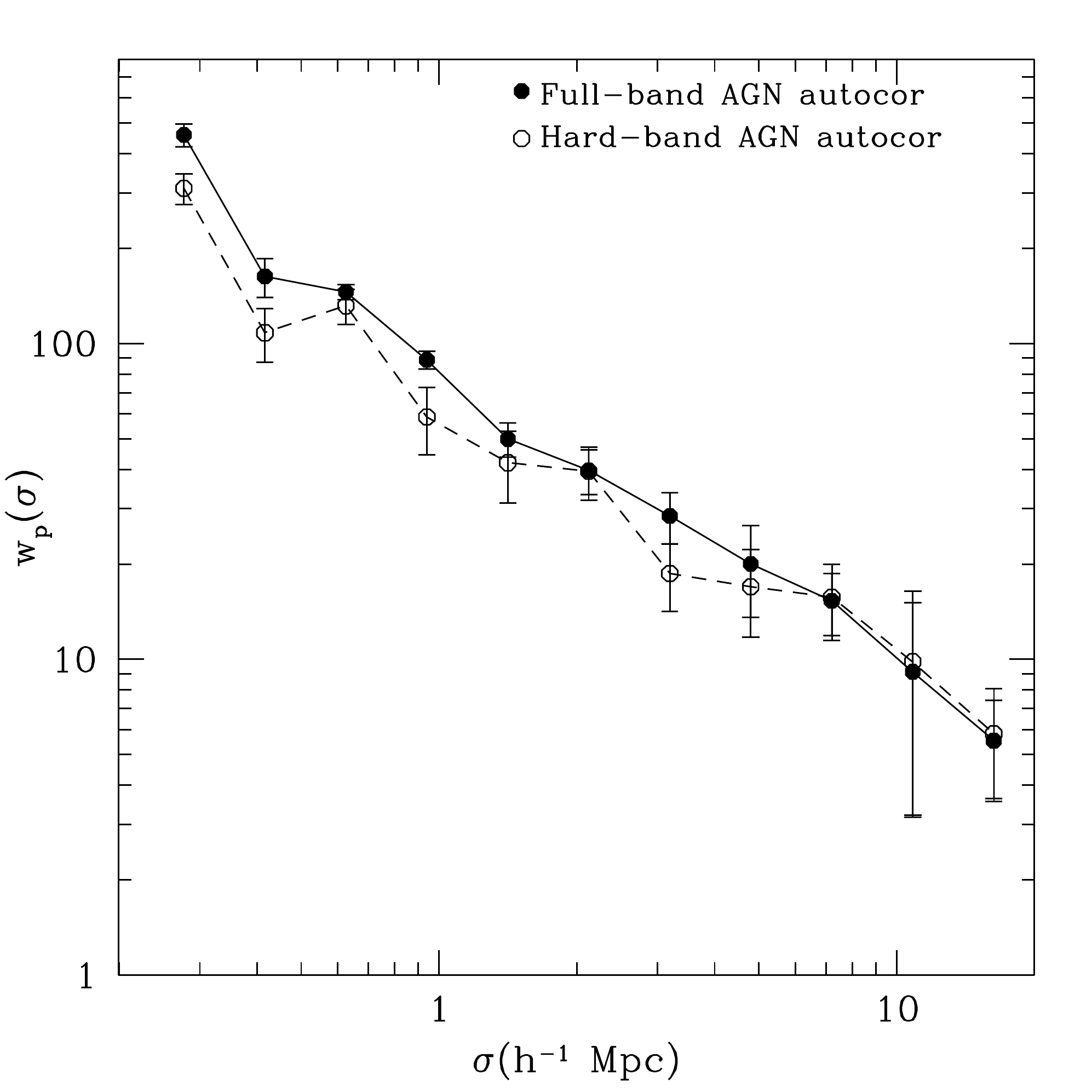}
\caption{Filled    circles   show    the   inferred    full-band   AGN
  auto-correlation  and  open   circles  the  inferred  hard-band  AGN
  auto-correlation. For clarity, open and filled circles are offset in the horizontal direction by $\delta \log \sigma = -0.01$ and +0.01, respectively.}
\label{fig:full_vs_hard}
\end{figure}

\section{Results}
Figure  $\ref{fig:fullband_agn}$  (left panel) compares  the  full-band  AGN-galaxy
cross-correlation    function,     $w_p(AG)$,    with    the    galaxy
auto-correlation  function, $w_p(GG)$.   Galaxies  are less  clustered
than  AGN at all  scales. The  infered AGN  projected auto-correlation
function,  $w_p(AA)$, (i.e.   equation  $\ref{eqn:inferred}$) is  also
plotted  in the  figure.  The  errors  in $w_p(AA)$  are estimated  by
adding  in quadrature  the uncertainties  of $w_p(AG)$  and $w_p(GG)$.
The right panel of Figure $\ref{fig:fullband_agn}$  shows the clustering  results for the
hard-band  AGN  sample. Within  the  errors  the  clustering of  those
sources  is similar to  the full-band  sample. This  is also  shown in
Tables   $\ref{table:x_rogamma}$   and  $\ref{table:rogamma}$,   which
present  the best-fit  estimates for  the $r_0$  and $\gamma$  for the
cross-correlation  and  the  auto-correlation functions  respectively,
assuming  a  power-law  form   for  $\xi$.   Scales  in  the  interval
$0.25-17$\,h$^{-1}$Mpc  are   used  to  fit  the   data.   The  errors
correspond to the 68th  percentile around the minimum $\chi ^2$.

\begin{figure*}
\begin{center}
\includegraphics[scale=0.57]{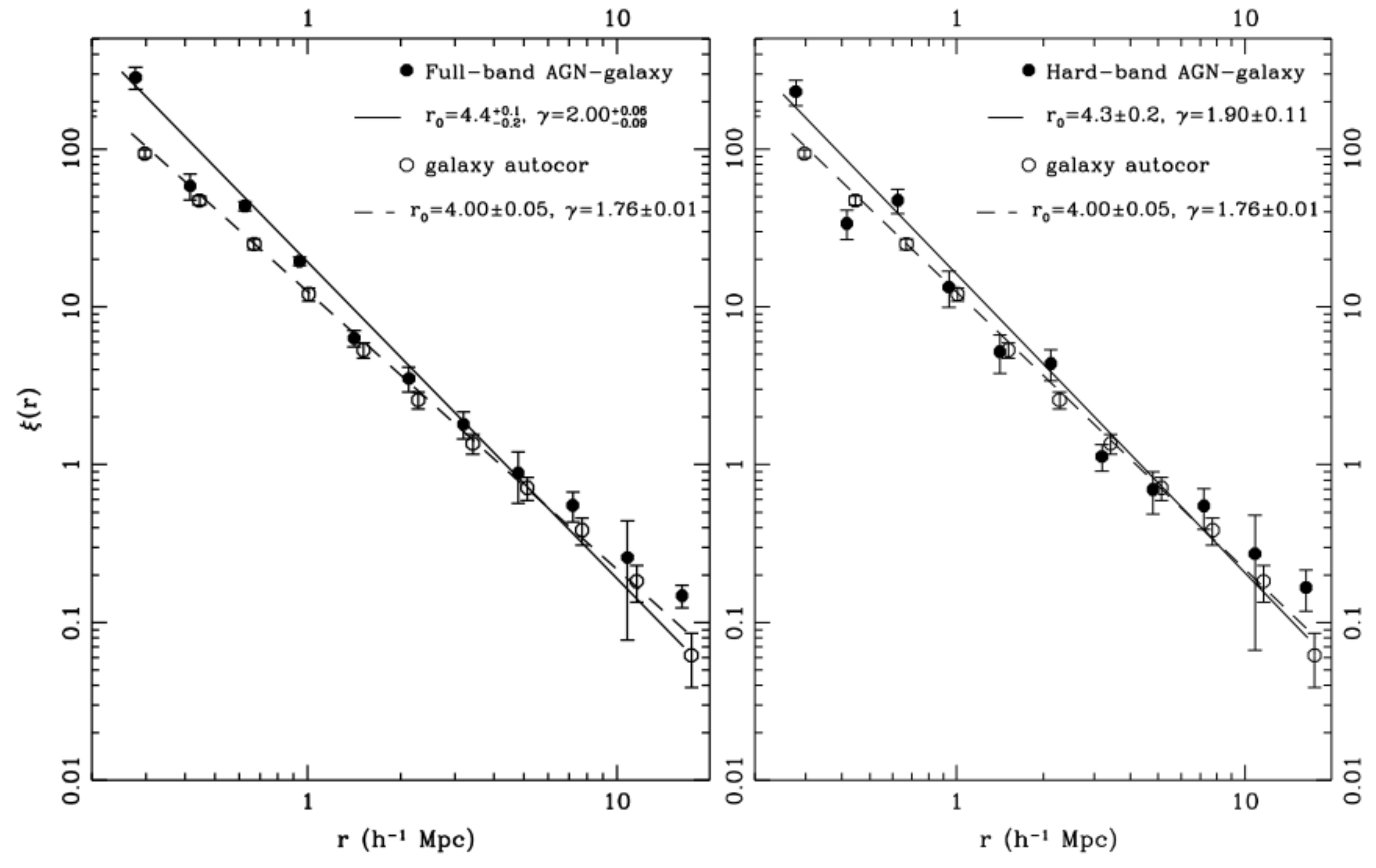}
\caption{Left panel: The   full-band  AGN-galaxy  cross-correlation   function  in
 real-space is shown by filled  circles. The solid line shows the fit
 on  scales  $0.25<r<17$h$^{-1}$Mpc.  Open  circles show  the  galaxy
 auto-correlation results and the fit is shown by the dashed line. Right panel: Same as in the left panel  but using the hard-band AGN sample. For clarity, open circles are offset in the horizontal direction by $\delta \log \sigma = +0.03$.}
\label{fig:xir_agn}
\end{center}
\end{figure*}

The  assumption  that the  real-space  correlation  function is  well
approximated    by   a    power   law    is   justified    by   Figure
$\ref{fig:xir_agn}$.  It  compares the inferred  $\xi(r)$ for
the  AGN/galaxy  cross-correlation  and  the  galaxy  auto-correlation
function without  making any assumptions on its  functional form (i.e.
equations  $\ref{eqn:xir_1}$, $\ref{eqn:xir_2}$)  with  power-law fits
using the $r_0$ and  $\gamma$ values of Tables $\ref{table:x_rogamma}$
and $\ref{table:rogamma}$.

We can therefore use the best-fit parameters for $r_0$ and $\gamma$ to
estimate    the    bias     of    AGN    and    galaxies    (equations
$\ref{eqn:xi20}-\ref{eqn:bs8}$).  The  results are presented  in Table
$\ref{table:bias}$,   where   b($\xi_{20}$)   and  b($\sigma_8$)   are
respectively the biases estimated from the first and the second method
described in section 3.2.  The errors are determined from the variance
across  the   six  subregions  of  the  surveyed   area.   Within  the
uncertainties  the  bias values,  calculated  from  the two  different
methods  are  consistent.   We  caution  that  the  b($\sigma_8$)  and
b($\xi_{20}$)  are  determined from  slightly  different scales.   The
former  is based  on the  $r_0$ and  $\gamma$ parameters  estimated by
fitting  the   projected  cross-correlation  function   on  scales  of
$0.25-17$\,h$^{-1}$Mpc.  The  latter is determined  by integrating the
$\xi(r)$ from  $1.0$ to 20\,h$^{-1}$\,Mpc. We  have confirmed however,
that this difference does not change the results and conclusions.

\begin{table*}
\caption{$r_0$ and $\gamma$ values for $\xi(r)$ (scales of $0.25-17h^{-1}$Mpc) for the cross-correlation measurements.}
\centering
\setlength{\tabcolsep}{4.0mm}
\begin{tabular}{lcccccc}
       \hline
$$ & {AGN (full )} &{Red AGN}&{Blue AGN}& {AGN (Hard)}&{Red galaxies}&{Blue galaxies}\\
       \hline
 $r_0$       & $4.4^{+0.1}_{-0.2}$ & $4.9\pm0.4$& $4.1\pm0.2$ & $4.3\pm0.2$  & $4.80\pm0.06$ & $3.60\pm0.05$ \\
$\gamma$ & $2.00^{+0.06}_{-0.09}$  & $1.98\pm0.10$ & $1.95\pm0.05$ & $1.90\pm0.11$ & $1.82\pm0.01$ & $1.71\pm0.01$ \\
       \hline
\label{table:x_rogamma}
\end{tabular}
\end{table*}

\begin{table*}
\caption{$r_0$ and $\gamma$ values for $\xi(r)$ (scales of $0.25-17h^{-1}$Mpc) for the galaxy auto-correlation and for the inferred auto-correlation function using equation $\ref{eqn:inferred}$.}
\centering
\setlength{\tabcolsep}{4.0mm}
\begin{tabular}{lccccccc}
       \hline
$$ & {AGN (full )} &{Red AGN}&{Blue AGN}& {AGN (Hard)}&{galaxies}&{Red galaxies}&{Blue galaxies}\\
       \hline
 $r_0$       & $5.0\pm0.5$ & $6.6\pm0.6$& $4.7\pm0.4$ & $4.8\pm0.6$  & $4.00\pm0.05$ & $5.00\pm0.10$&$2.90\pm0.05$ \\
$\gamma$ & $2.00\pm0.11$  & $1.82\pm0.19$ & $2.12\pm0.16$ & $2.02\pm0.18$ & $1.76\pm0.01$ & $1.90\pm0.02$&$1.58\pm0.01$  \\
       \hline
\label{table:rogamma}
\end{tabular}
\end{table*}

\begin{figure}
\begin{center}
\includegraphics[scale=0.43]{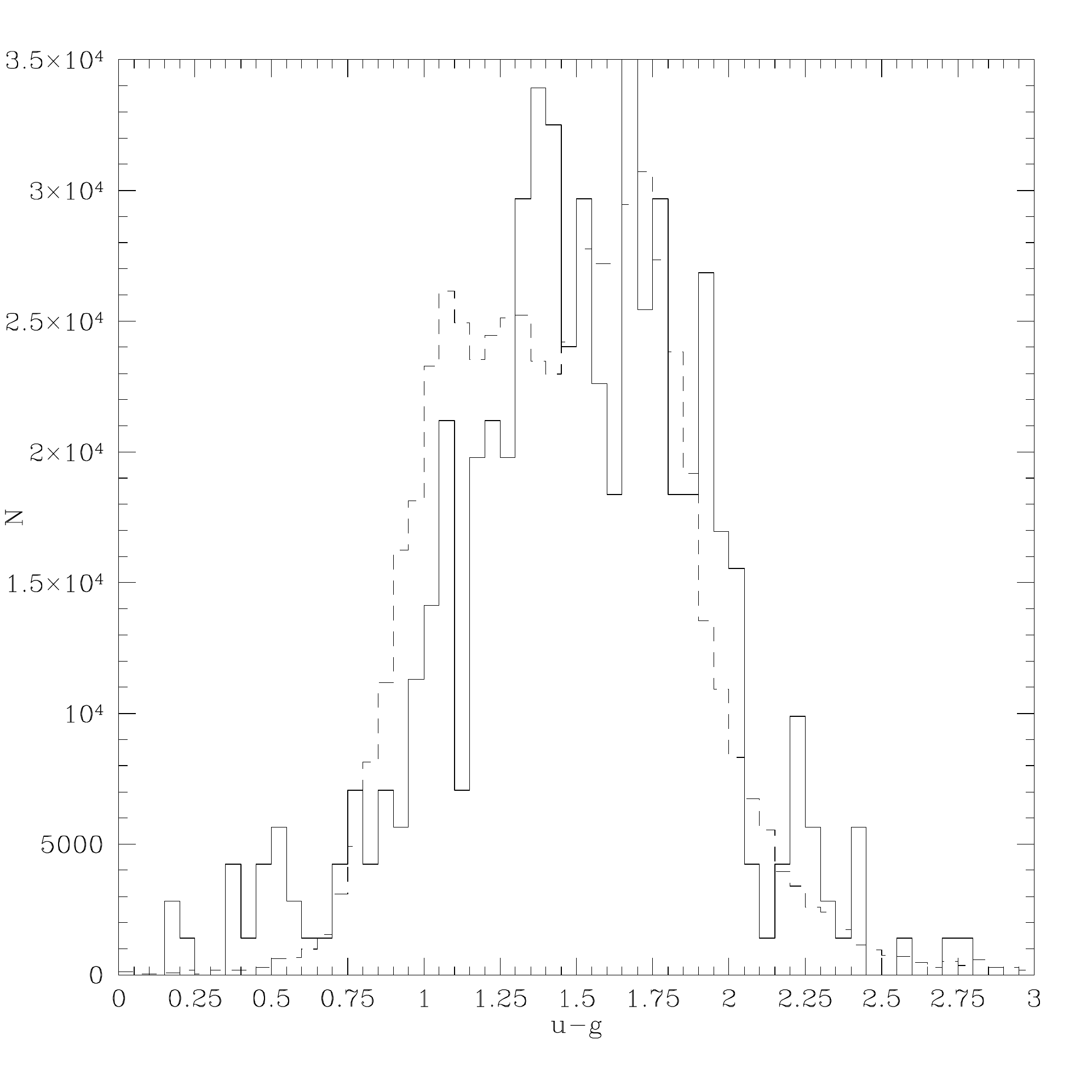}
\caption{The $^{0.1}u-g$ histograms for the  galaxies (solid line) and the AGN
  (dashed line).  For  clarity, the AGN histogram has  been scaled, by
  the  ratio  $N_G/N_A$, where  $N_G$  and  $N_A$  are the  number  of
  galaxies and AGN, respectively.}
\label{fig:ug_dist}
\end{center}
\end{figure}

Next,  we explore  variations  in  the clustering  of  X-ray AGN  with
rest-frame  optical  colour. This  is  motivated  by previous  studies
suggesting that AGN  in red cloud hosts are  more clustered than those
in blue cloud galaxies (e.g. Coil  et al. 2009).  For this exercise we
use  the  full-band  AGN  sample  because  of  its  larger  size.
Red/blue galaxies/AGN are  separated by their $^{0.1}(u-g)$ rest-frame
colour, i.e.   the difference between  the absolute magnitudes  of the
source in the $^{0.1}u$, $^{0.1}g$  bands, which are the SDSS $u$, $g$
filters  shifted  to $z=0.1$.   The  calculation  of $^{0.1}(u-g)$  is
carried out using the {\sc  kcorrect} version 4.2 routines (Blanton \&
Roweis  2007).   The  distribution   of  X-ray  AGN  and  galaxies  in
$^{0.1}(u-g)$  colour is  plotted in  Figure  $\ref{fig:ug_dist}$. The
division  between   red  and   blue  galaxies  and   AGN  is   set  to
$^{0.1}(u-g)=1.5$ (Blanton 2006).  There  are 325,510 galaxies and 154
X-ray  AGN with  colours bluer  than  that cut  (blue subsamples)  and
266,508  galaxies  and  143  X-ray  AGN  with  $^{0.1}(u-g)>1.5$  (red
subsamples).

\begin{figure}
\begin{center}
\includegraphics[scale=0.287]{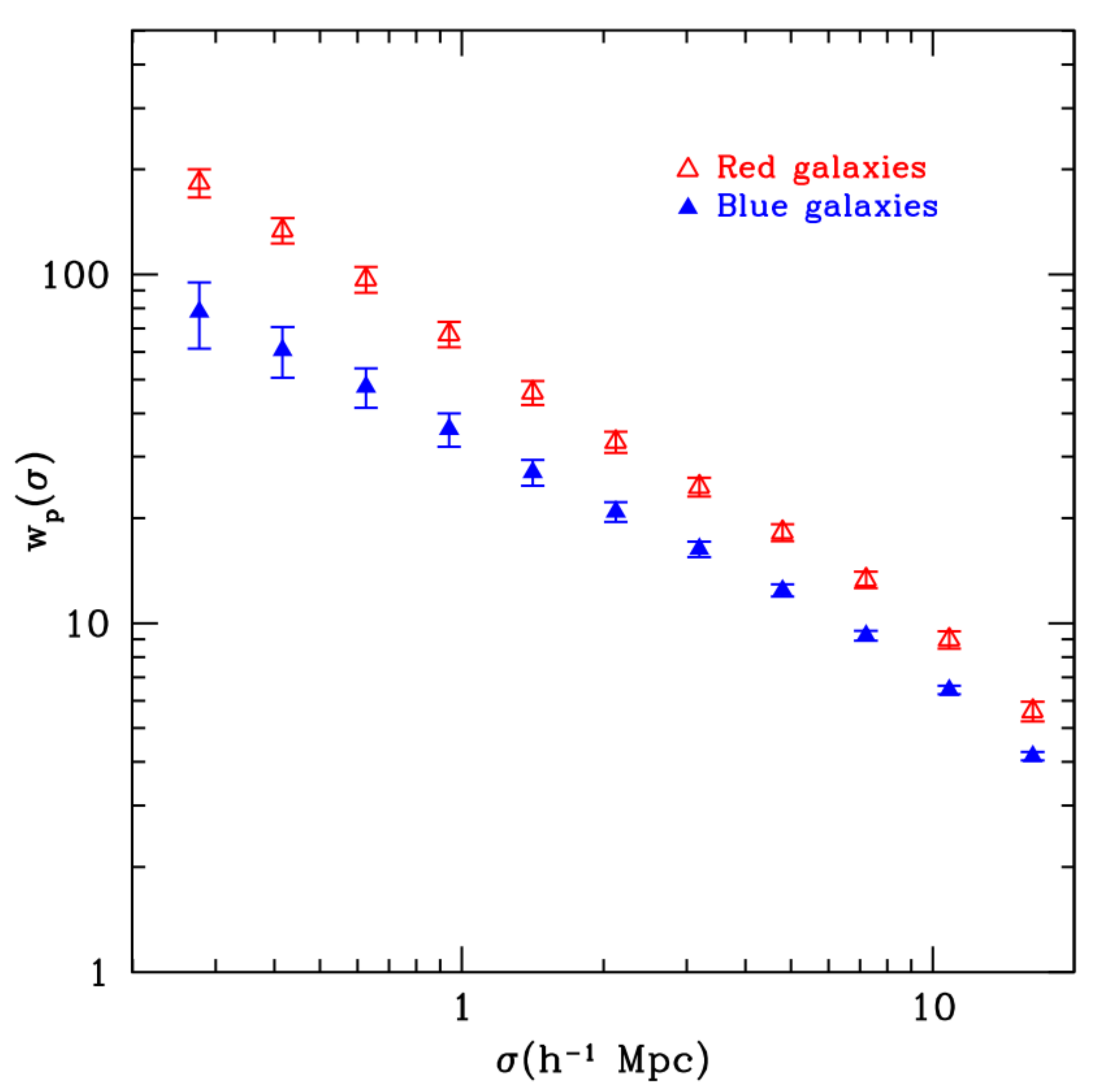}
\caption{Cross-correlation functions of blue galaxies ($^{0.1}(u-g)<1.5$) with
  the galaxy sample (filled  triangles) and red galaxies with the
  galaxy sample (open  triangles). Red galaxies cluster more than blue galaxies.}
\label{fig:gal_colors}
\end{center}
\end{figure}

\begin{figure}
\begin{center}
\includegraphics[scale=0.45]{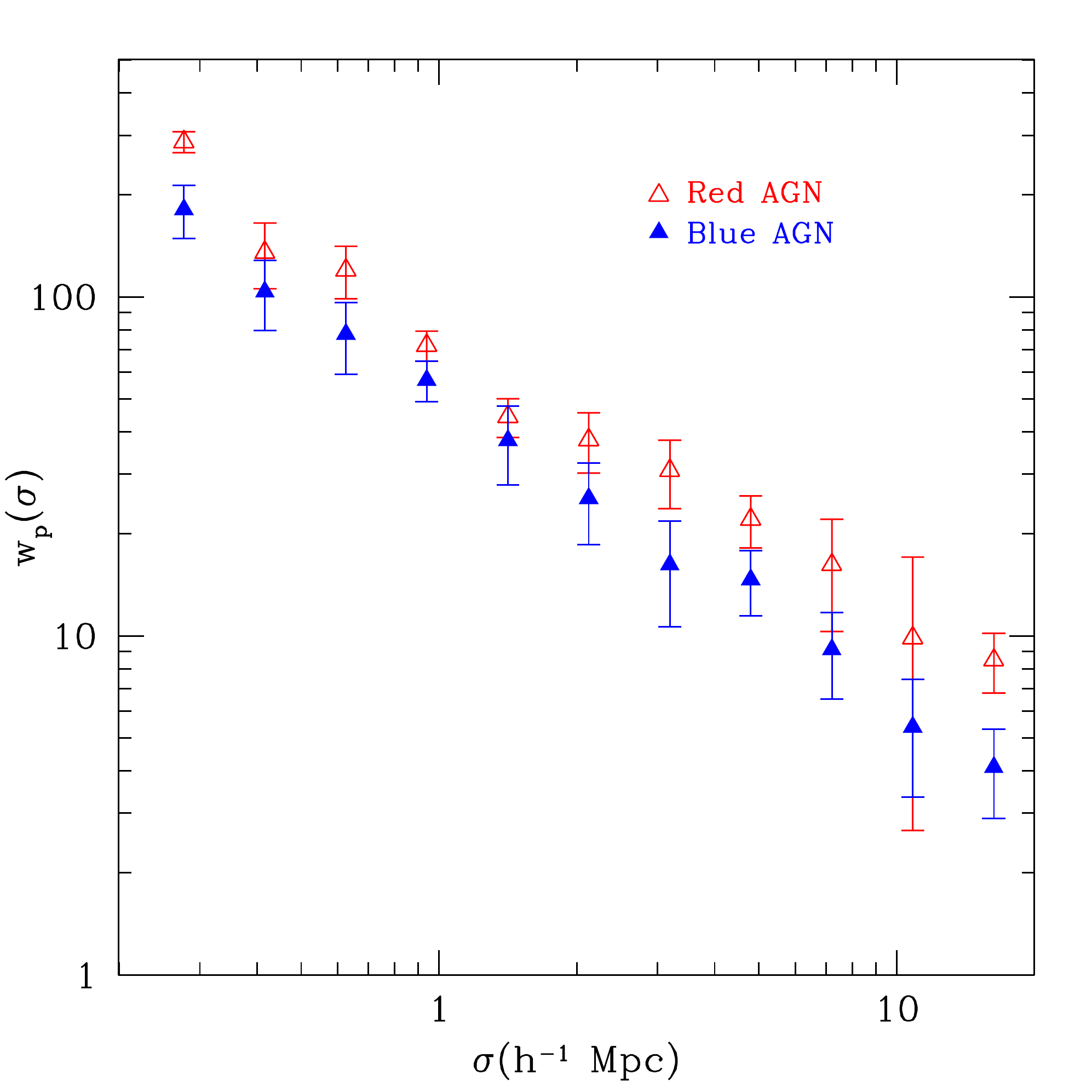}
\caption{Blue   ($^{0.1}(u-g)<1.5$)   AGN cross-correlation   function
  with galaxies (filled triangles) versus the red-AGN/galaxy cross-correlation  function (open triangles). Red AGN cluster more than blue AGN. For clarity, open and filled triangles are offset in the horizontal direction by $\delta \log \sigma = -0.01$ and +0.01, respectively.}
\label{fig:agn_colors}
\end{center}
\end{figure}

Figure $\ref{fig:gal_colors}$ plots the projected cross-correlation of
red and blue galaxies with  the full galaxy sample. The relevant $r_0$
and  $\gamma$ best-fit  values, bias  parameters and  dark  matter halo
masses    are    shown    in    Tables    $\ref{table:rogamma}$    and
$\ref{table:bias}$.  As expected red  galaxies are more clustered than
blue  ones   at  all  scales,  in  agreement   with  previous  studies
(e.g. Madgwick  et al.  2003, Zehavi  et al.  2005, Coil  et al. 2008,
Hickox  et al.   2009).  The  projected cross-correlation  function of
red/blue AGN  with the  overall galaxy population  is shown  in Figure
$\ref{fig:agn_colors}$. The clustering results are presented in Tables
$\ref{table:rogamma}$  and $\ref{table:bias}$.   X-ray AGN  follow the
same pattern  with galaxies,  i.e. active SMBH  in red hosts  are more
clustered than those in blue galaxies.

This is further demonstrated in Table $\ref{table:relativebias}$ which
presents  the relative bias  of different  subsamples, defined  as the
ratio  between  the   $w_p(\sigma)$  measurements  of  the  subsample,
$b_{rel}=\sqrt{{w_p(\sigma)_1}/{w_p(\sigma)_2}}$ (Coil  et al.  2007),
integrated     over     two     scales,    $0.25-8$h$^{-1}$Mpc     and
$1-8$h$^{-1}$Mpc. The errors on  these measurements are estimated from
the variance  of the  relative bias across  the six subregions  of the
survey  area (Section  3.1).  This  table confirms  that  red AGN  and
galaxies cluster more than blue AGN and galaxies, respectively.  Also,
the overall X-ray AGN  population has similar clustering properties as
red galaxies but is more clustered than blue ones.

We also investigate the dependence of the clustering on luminosity and
X-ray obscuration by  splitting the sample into two  nearly equal size
groups  at  $\log  L_X=41.8$  (erg/s) and  $\log  N_H=22$  (cm$^{-2}$)
respectively. We do not find any statistically significant trends of
the AGN clustering with luminosity or X-ray hardness.  This null
result maybe be because of the relatively narrow luminosity and
obscuration baselines of our low  redshift X-ray AGN   
sample. Cappelluti  et al. (2010) for example, find a higher
clustering for type I AGN compared to type IIs. Their sample is drawn
from the  SWIFT-BAT AGN  catalogue, which is sensitive to heavily
obscured systems. Also, Krumpe et al. (2010) find evidence 
for higher clustering  only for very bright AGN,  $\rm L_X >10^{44} \, erg
s^{-1}$. The low redshift subset  of the XMM/SDSS survey does not have
the volume to detect such powerful sources.

Finally, we  calculate the masses of  the dark matter  halos that host
AGN and galaxies, as described in Section 3.2. The results appear in Table
$\ref{table:bias}$. We do not estimate a lower limit for the mass of
the halos that host blue galaxies. This is because the $1\sigma$ lower
limit of  the bias of those  sources is $b=0.69$ and  the minimum bias
value derived from Sheth 
et  al.  (2001) is  $b=0.72$.   The  calculations  show that  moderate
luminosity  AGN at  $z\approx0.1$  reside in  dark  matter halos  with
typical mass $M_{DMH} \approx 10^{13}\rm \, h^{-1} \, M_{\sun}$.

The AGN bias and DMH mass estimates presented in Table 3 are robust to
the details of the adopted methodology. Cosmic variance
is a concern as in the cross-correlation function estimation we use
all SDSS galaxies, not just those overlapping with the XMM
pointings. Therefore the cosmological volumes of the AGN and galaxy
samples are different. We repeat the analysis for the full-band
selected AGN sample using only those SDSS galaxies that lie within the
XMM fields of our serendipitous survey. This excerise yields the same
AGN bias as in Table 3. Additionally, the exclusion of XMM pointings 
targeting clusters (see section 2) might appear arbitrary and
subjective. We include AGN in those fields and cross-correlate them
with the SDSS galaxies in the same cosmological volume, i.e. those
overlaping with the  XMM pointings of the XMM/SDSS survey. The infered
full-band AGN bias is 10 per cent lower compared to that listed
in Table 3, i.e. within the estimated errors.

\begin{table*}
\caption{Bias values for AGN and galaxy samples  using the two approaches discussed in the text. Values marked with $^*$ are calculated for $\sigma_8=0.84$. The $\xi_{20}$ values are from the cross-correlation measurements (except for galaxies). The M$_{DMH}$ are based on b$(\sigma_8=0.8)$.}
\centering
\setlength{\tabcolsep}{3.0mm}
\begin{tabular}{lccccccc}
       \hline
$$ & {AGN (full)} &{Red AGN}&{Blue AGN}& {AGN (hard)}&{galaxies}&{Red galaxies}&{Blue galaxies}\\
       \hline
       \\
$\xi_{20}$ & $0.22^{+0.03}_{-0.05}$ & $0.26^{+0.02}_{-0.05}$& $0.18^{+0.04}_{-0.03}$ & $0.21^{+0.03}_{-0.03}$  & $0.17^{+0.02}_{-0.01}$ & $0.23^{+0.03}_{-0.02}$ & $0.15^{+0.01}_{-0.01}$ \\[3pt]
$\sigma_{8,AGN}$ & $0.94\pm0.10$  & $1.16\pm0.16$ & $0.92\pm0.11$ & $0.91\pm0.13$ & $0.73\pm0.01$ & $0.91\pm0.01$&$0.57\pm0.01$  \\[3pt]
b($\xi_{20})$ & $1.30^{+0.20}_{-0.27}$  & $1.52^{+0.17}_{-0.28}$ & $1.07^{+0.24}_{-0.17}$ & $1.25^{+0.20}_{-0.12}$ & $0.93^{+0.05}_{-0.03}$ & $1.37^{+0.20}_{-0.12}$ & $0.89^{+0.10}_{-0.07}$  \\[3pt]
b$^*(\xi_{20})$ & $1.26^{+0.18}_{-0.25}$  & $1.47^{+0.15}_{-0.27}$ & $1.05^{+0.22}_{-0.15}$ & $1.15^{+0.13}_{-0.16}$ & $0.89^{+0.05}_{-0.03}$ & $1.32^{+0.19}_{-0.11}$&$0.86^{+0.09}_{-0.06}$  \\[3pt]
b($\sigma_8)$ & $1.23^{+0.12}_{-0.17}$  & $1.52^{+0.12}_{-0.10}$ & $1.20^{+0.10}_{-0.16}$ & $1.20^{+0.15}_{-0.12}$ & $0.96^{+0.05}_{-0.05}$ & $1.20^{+0.11}_{-0.06}$&$0.75^{+0.06}_{-0.06}$  \\[3pt]
b$^*(\sigma_8)$ & $1.17^{+0.11}_{-0.15}$  & $1.45^{+0.11}_{-0.08}$ & $1.14^{+0.09}_{-0.15}$ & $1.14^{+0.14}_{-0.11}$ & $0.92^{+0.04}_{-0.04}$ & $1.14^{+0.10}_{-0.05}$&$0.71^{+0.05}_{-0.05}$  \\[3pt]
log$\left(\frac{M_{DMH}}{h^{-1}M_{\sun}}\right)$& $12.93^{+0.22}_{-0.29}$  & $13.30^{+0.26}_{-0.44}$ &  $12.88^{+0.20}_{-0.40}$ & $12.88^{+0.27}_{-0.32}$ & $12.18^{+0.18}_{-0.24}$ & $12.88^{+0.20}_{-0.14}$&$10.46^{+0.80}_{}$  \\[3pt]
       \hline
\label{table:bias}
\end{tabular}
\end{table*}

\section{Comparison with Previous Studies}

In this section our bias  estimates for X-ray AGN at $z\approx0.1$ are
compared  with previous  studies  that select  active  SMBH either  at
X-rays or UV/optical. To  facilitate the direct comparison and present
the results from  different papers in a uniform  manner we re-estimate
the bias from each study  using the quoted $r_0$ and $\gamma$ values
based on the $\sigma_8$ methodology  of section 3.1. Studies that have
measured  only  the redshift-space  correlation  function  and do  not
provide  real-space $r_0$ and  $\gamma$ values  are excluded  from the
analysis. Moreover, angular auto-correlation function studies are
not included  in our  compilation as the  deprojection of  the angular
signal to  3-dimensions using the  Limber's formula (Limber  1953) may
introduce systematic  uncertainties.

\begin{figure*}
\begin{center}
\includegraphics[scale=0.9]{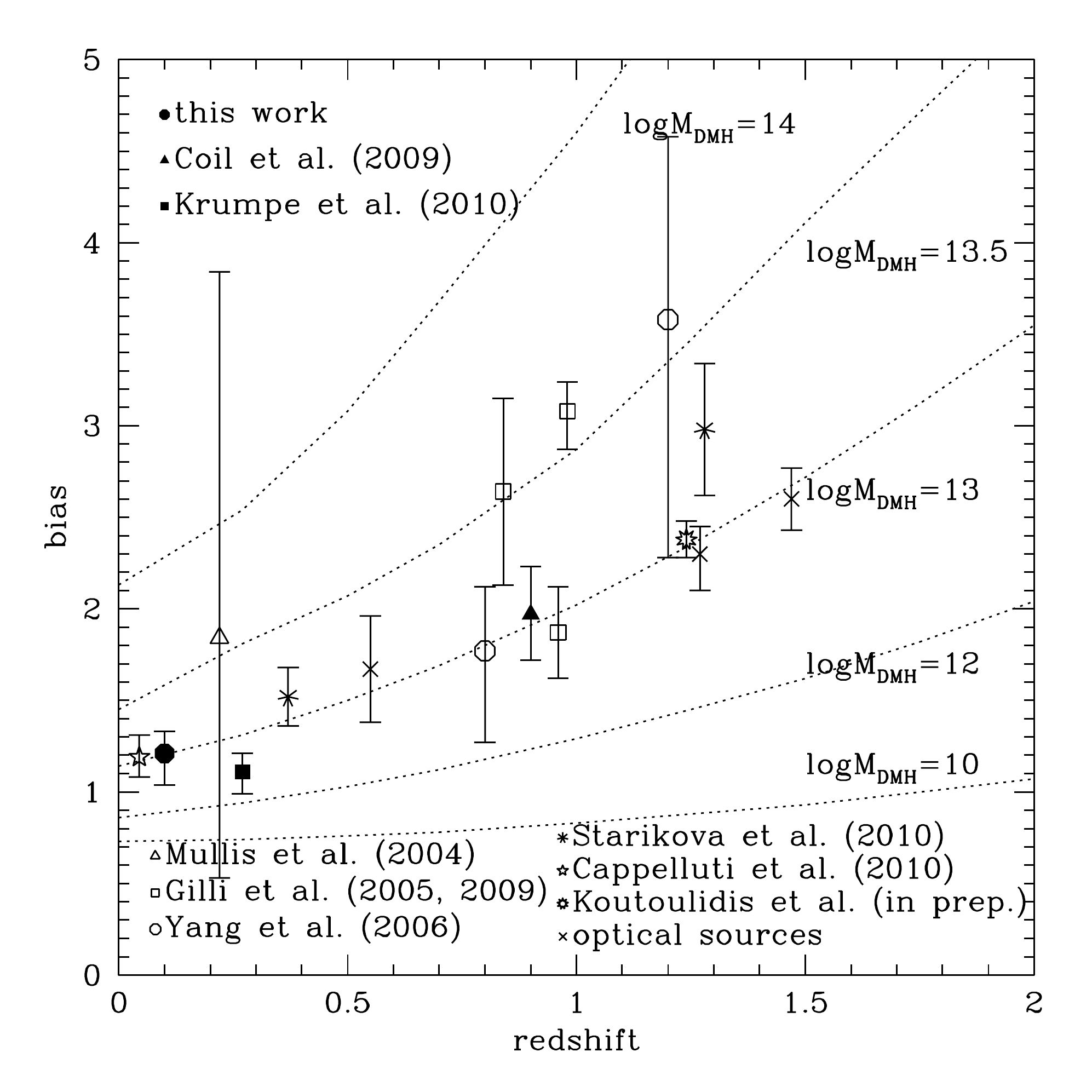}
\caption{Comparison of  our bias  estimation (filled circle)  to other
  studies.  Crosses present the  estimated biases for optical samples.
  From low redshift to high  redshift, Mountrichas et al. (2008), Ross
  et  al. (2009)  and Ivashchenko  et al.   (2010).  The  dotted lines
  present the expected $b(z)$ of  dark matter halo masses. Dark matter
  halos  have been estimated  as described  in the  text (in  units of
  h$^{-1}$Mpc).}
\label{fig:bias_dmh}
\end{center}
\end{figure*}

Table  $\ref{table:studies}$  presents  the  studies  used  for  these
measurements  and the most  important properties  of the  samples they
used. The errors  on b($\sigma_8$) are calculated based  on the errors
on  $r_0$,   $\gamma$.  The  bias  parameter  is   compared  with  our
measurement in Figure $\ref{fig:bias_dmh}$ and the expected evolution
of the bias for different dark matter halo masses. Below we comment
on selected datapoints appearing in Table $\ref{table:studies}$.

From the high redshift  samples plotted in Figure $\ref{fig:bias_dmh}$
the Coil  et al. (2009) datapoint  is the most  appropriate to compare
with our measurement of the  bias at $z\approx0.1$.  They select X-ray
AGN at  $z\approx1$ at rest-frame energies of  about 1-14\,keV similar
to those  used here. The luminosity  range of their AGN  is similar to
our low redshift sample.  They also use the cross-correlation function
with  galaxies  to  improve  on  the statistics  and  minimise  sample
variance effects in  their calculations.  The mass of  the dark matter
halos that host  these high redshift X-ray AGN,  as estimated based on
the quoted $r_0$ and $\gamma$ values by Coil et al. (2009), is in
good agreement with our measurements at low redshift.  This shows
that moderate luminosity X-ray AGN live in dark matter haloes of
similar mass, $\sim 10^{13} \, h^{-1}\,M_{\sun}$, at all redshifts out
$z\approx1$.

\begin{figure}
\begin{center}
\includegraphics[scale=0.45]{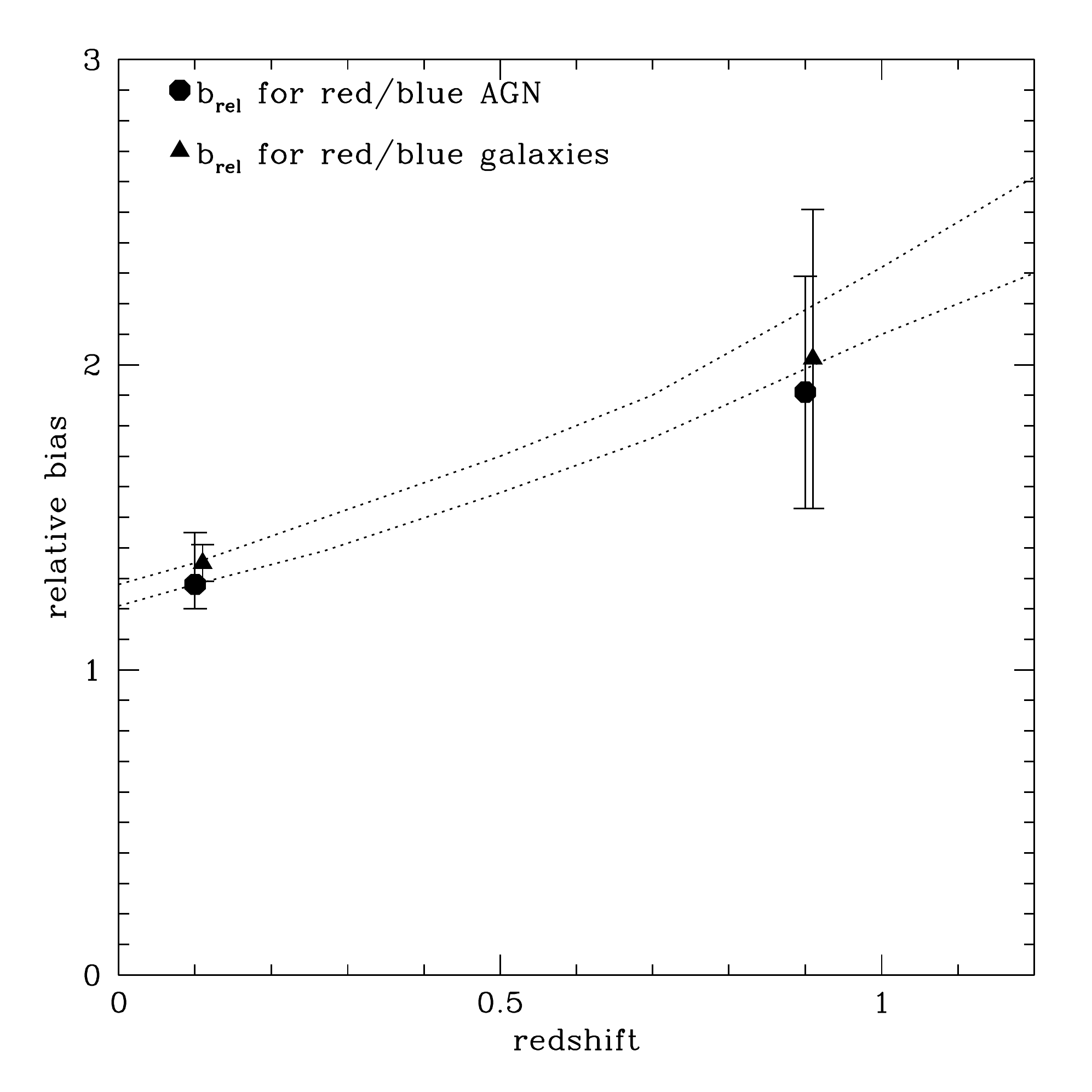}
\caption{Relative bias for red/blue AGN (filled circle at $z=0.1$) and red/blue galaxies (filled triangle). Triangles are offset by 0.01 for clarity. Dotted lines show how the (relative) bias evolves with redshift.  At $z=0.9$ the $b_{rel}$ of Coil et al. (2009) are plotted. The higher relative bias values at $z=0.9$ are in agreement with those at $z=0.1$, given the redshift evolution of the relative bias.}
\label{fig:relative_bias_dmh}
\end{center}
\end{figure}

In agreement  with Coil  et al. (2009)  we also  find that AGN  in red
hosts are more clustered than those in blue galaxies. However, they find that  the difference in AGN clustering due  to the host color
is more apparent for their $z\sim1$ AGN than in our sample at $z=0.1$.
The  relative bias between red and  blue AGN at $z\approx1$, $b_{rel}\sim1.9$
(see  Table  2 of Coil et al. 2009)  compared to  $b_{rel}\sim1.3$ at  $z\approx0.1$
(Table  $\ref{table:relativebias}$).   Similarly,  our  relative  bias
estimation for  the red/blue  galaxies is $b_{rel}\sim1.35$.   Coil et
al. (2009) do not give a value  for this measurement, but using their quoted
relative bias values  for red/blue AGN, red AGN/red  galaxies and blue
AGN/blue  galaxies  we  infer  a  red/blue  galaxy  relative  bias  of
$b_{rel}=2.02\pm0.49$  at $z\approx1$.   This apparent  discrepancy is
because of  the differential  evolution with redshift  of the  bias of
dark  matter  halos of  different  masses.   This  is demonstrated  in
Fig. $\ref{fig:relative_bias_dmh}$  which shows how  the relative bias
of red/blue AGN and galaxies evolves with redshift.

Gilli et al. (2005) and (2009) measured the projected auto-correlation
function of X-ray AGN in  the Chandra Deep Field South (CDFS), Chandra
Deep Field North (CDFN) and the  COSMOS surveys.  AGN in the CDFs have
luminosities comparable to those of  Coil et al.  (2009), although the
COSMOS sample  is more  X-ray luminous, because  of the  large angular
size ($\rm 2\, deg^2$) and shallower depth of the X-ray survey in that
field.  Gilli  et al.   (2005, 2009) find  that sample variance  has a
strong effect on their results,  even in the wide-angle COSMOS sample,
resulting  in   large  values  for  the  clustering   length  and  the
corresponding  dark matter  halo masses,  $\rm  \log M_{DMH}\sim13.5\,
h^{-1} \,  M_{\sun}$.  Narrow peaks  in the AGN  redshift distribution
have  a significant  contribution  to the  clustering  signal and  are
primarily   responsible  for   the  high   $\rm   M_{DMH}$  estimates.
Preferentially removing galaxies in those peaks decreases the AGN bias
in the  CDFS and the COSMOS to  values comparable to those  of Coil et
al.   (2009) and  to our  estimates. However,  Marulli et  al.  (2009)
argue  that sample variance  cannot account  for the  large clustering
length  in the  CDFS.  They  use  semi-analytic models  to follow  the
cosmological  evolution of AGN  and produce  mock AGN  catalogues with
selection functions  similar to the  CDFN and the CDFS.   Although the
clustering of AGN  in their simulations is in  good agreement with the
observations in the CDFN (Gilli  et al.  2005) they cannot account for
the large clustering signal in the CDFS within $2.0-2.5\sigma$.  These
results highlight  the importance of developing  methods which account
for  sample  variance  (e.g.   cross-correlation  function,  see  also
Allevato et al. 2011).

In  addition  to  X-ray  selected AGN,  Figure  $\ref{fig:bias_dmh}$
(crosses)  and Table  $\ref{table:studies}$ also  include measurements
for the  bias of  UV/optically selected QSOs  from Mountrichas  et al.
(2009), Ross et al. (2009)  and Ivashchenko et al.  (2010). The latter
study  measured  the auto-correlation  function  of $\sim50,000$  SDSS
quasars  at $\bar{z}=1.47$.   We estimate  that their  quasar  bias is
$b=2.06\pm0.10$    which   is    higher   than    their   calculation,
$b=1.44\pm0.22$.  This discrepancy is  likely because of the different
methodology  they  follow  for  the  bias estimation.   They  use  the
redshift-space  and real-space  measurements to  calculate  the infall
parameter ($\beta$;  see their  eqn 7) and  based on  this calculation
they estimate the AGN bias ($\beta=f(\Omega_m,z)/b$; see their section
4.1  for  more  details).   Ross  et  al.   (2009)  used  $\sim30,000$
spectroscopic   SDSS  quasars,  over   the  redshift   range  $0.3\leq
z\leq2.2$.     We    estimate   a    bias    for    their   QSOs    of
$b=2.30^{+0.15}_{-0.20}$   (Table  $\ref{table:studies}$).    This  is
higher than $b=2.06\pm0.03$ that they find using a methodology similar
to that of Ivashchenko et al.  (2010). The discrepancy between the two
numbers is  likely because of  the different approaches used  to infer
the QSO bias.

Figure $\ref{fig:bias_dmh}$ shows that in a redshift range $0<z<2$ the
AGN bias datapoints  cluster around the curve that  correponds to dark
matter  halos with  mass $\rm  M_{DMH}\sim10^{13} \,h^{-1}\,M_{\sun}$.
This seems to  be the case for both moderate  luminosity X-ray AGN and
powerful UV/optical  quasars as  presented above. We  caution however,
that many optical  studies (Croom et al.  2005, Myers  et al. 2007, da
$\hat{A}$ngela et  al.  2008)  find lower masses  for the  dark matter
halos   of   powerful   UV   bright   QSO,   $\rm   M_{DMH}   \sim   5
\times10^{12}\,h^{-1}  \, M_{\sun}$.  The  results from  those studies
are not shown in Figure $\ref{fig:bias_dmh}$ for the reasons discussed
at the beginning of the section.

\begin{table*}
\caption{Results for the relative bias.}
\centering
\setlength{\tabcolsep}{4.0mm}
\begin{tabular}{lcc}
       \hline
$Sample$ & Relative bias & Relative bias\\
& $0.25<\sigma<8h^{-1}Mpc$ & $1.0<\sigma<8h^{-1}Mpc$\\[1pt]
       \hline
       \\
AGN full/hard       & $1.05^{+0.05}_{-0.08}$ & $1.03^{+0.04}_{-0.07}$ \\[3pt]
AGN red/blue  & $1.28^{+0.17}_{-0.08}$ & $1.25^{+0.15}_{-0.10}$ \\[3pt]
galaxies red/blue  & $1.35^{+0.06}_{-0.06}$ & $1.29^{+0.04}_{-0.04}$ \\[3pt]
red AGN/red galaxies  & $1.09^{+0.05}_{-0.03}$ & $1.07^{+0.04}_{-0.03}$ \\[3pt]
blue AGN/blue galaxies  & $1.15^{+0.04}_{-0.07}$ & $1.07^{+0.05}_{-0.05}$ \\[3pt]
(full) AGN/red galaxies &   $0.91^{+0.03}_{-0.06}$ & $0.95^{+0.02}_{-0.06}$ \\[3pt]
(hard) AGN/red galaxies &   $0.91^{+0.04}_{-0.05}$ & $0.91^{+0.02}_{-0.06}$ \\[3pt]
(full) AGN/blue galaxies &   $1.37^{+0.08}_{-0.08}$ & $1.30^{+0.06}_{-0.06}$ \\[3pt]
(hard) AGN/blue galaxies  & $1.32^{+0.08}_{-0.08}$ & $1.25^{+0.05}_{-0.05}$ \\[3pt]
       \hline
\label{table:relativebias}
\end{tabular}
\end{table*}

\begin{table*}
\caption{Clustering  measurements for  X-ray and
  UV bright  AGN from the literature.  Columns are:  (1) reference to the AGN sample. A star marks UV/optical selected QSOs; (2) methodology used to determine the clustering, i.e. auto-correlation or cross-correlation; (3) name of the survey that the AGN sample was selected from; (4) number of sources used; (5) scales used each study to determine the clustering amplitude and slope; (6) best-fit clustering amplitude  measured in each study; (7) best-fit clustering slope measured in each study; (8) mean redshift of the sample; (9) the bias  values  re-calculated using  the $\sigma_8$ methodology  discussed in  the text.}       
\centering
\setlength{\tabcolsep}{1.2mm}
\scalebox{1.0}{
\begin{tabular}{ccccccccc}
       \hline  
       Study & Methodology  & Sample  & no  & scales&  $r_0$ &
       $\gamma$ &  z & b$(\sigma_8)$ \\ 
        &  & & of objects& (h$^{-1}$Mpc)  & (h$^{-1}$Mpc) &  & &   \\
        (1) & (2) & (3) & (4) & (5) & (6) & (7) & (8) & (9)   \\
       \hline 
       \\
       This  Work (full)  & cross-cor  & XMM/SDSS  &  297 &
       $0.25-17$ & $5.0\pm0.5$ & $2.00\pm0.11$ &  0.10 &  $1.23^{+0.12}_{-0.17}$\\[5pt]
       This Work  (hard) &
       cross-cor  &  XMM/SDSS  &  175  &  $0.25-17$  &  $4.8\pm0.6$  &
       $2.02\pm0.18$  &   0.10  &
       $1.20^{+0.15}_{-0.12}$\\[5pt]
       Coil et al. (2009) & cross-cor & AEGIS
       & 113  & $0.1-8$ &  $5.95\pm0.90$ & $1.66\pm0.22$  & 0.90 &  $1.97^{+0.26}_{-0.25}$\\[5pt] 
       Krumpe et
       al.   (2010)   &  cross-cor  &   RASS  &  1552  &   $0.3-15$  &
       $4.28^{+0.44}_{-0.54}$  &  $1.67^{+0.13}_{-0.12}$ &  0.27  & $1.11^{+0.10}_{-0.12}$\\[5pt]
        Mullis et  al.    (2004)&  auto-cor   &  NEP  &   219  &   $5.0-60$  &
       $7.5^{+2.7}_{-4.2}$   &  $1.85^{+1.90}_{-0.80}$  & $0.22$ & $1.84^{+2.00}_{-1.31}$ \\[5pt] 
       Gilli et al. (2005) & auto-cor &  CDFS & 97 & $0.2-10$ & $10.3\pm1.7$
       &  $1.33\pm0.14$   & 0.84  &
       $2.64\pm0.51$\\ [5pt]
       Gilli et  al. (2005) & auto-cor &  CDFN & 160 &
       $0.2-10$ & $5.5\pm0.6$ &  $1.50\pm0.12$ &  0.96 &  $1.87\pm0.25$\\ [5pt] 
       Gilli  et  al.  (2009)  &
       auto-cor &  COSMOS & 538 & $0.3-40$  & $8.65^{+0.41}_{-0.48}$ &
       $1.88^{+0.06}_{-0.07}$  & 0.98
       & $3.08^{+0.16}_{-0.21}$\\ [5pt]
       Yang et al. (2006) & auto-cor & CDFN
       & 252 & $0.2-15$ & $5.8^{+1.0}_{-1.5}$ & $1.38^{+0.12}_{-0.14}$
       &        0.8     &
       $1.77^{+0.35}_{-0.50}$\\ [5pt]
       Yang et al. (2006) & auto-cor & CLASXS
       &  233  &  $1.0-30$   &  $8.1^{+1.2}_{-2.2}$  &  $2.1\pm0.5$   &     1.2      &
       $3.58^{+1.00}_{-1.30}$\\ [5pt]
       Starikova  et al. (2010)  & auto-cor &
       Bootes  &  1282 &  $0.5-20$  &  $5.4\pm0.5$  & $1.97\pm0.09$  &     0.37     &
       $1.52\pm0.16$\\ [5pt]
       Starikova  et al. (2010) & auto-cor  & Bootes &
       1282  & $0.5-20$  &  $7.0\pm0.8$ &  $1.97\pm0.09$ & 1.28 & $2.98\pm0.36$\\  [5pt]
       Cappelluti et al.
       (2010)   &    auto-cor   &   BAT   &   199    &   $0.2-200$   &
       $5.56^{+0.49}_{-0.43}$  &  $1.64^{+0.08}_{-0.07}$ &        0.045        &
       $1.19^{+0.12}_{-0.11}$\\ [5pt]
       Koutoulidis et al. (in prep.) & auto-cor & CDFs/ECDFs & 1492 & $0.2-20$ & $6.1\pm0.2$ & 1.8 (fixed) & 1.24 & $2.4\pm0.1$\\ & & COSMOS/AEGIS &&&&&&\\[5pt]
       Mountrichas   et  al.   (2008)$^*$  &
       cross-cor  &   2SLAQ  &  694  &  $5.0-25.0$   &  $6.7\pm0.6$  &
       $1.75\pm0.10$ & 0.55 & $1.67\pm0.29$\\ [5pt]
        Ross et
       al.   (2009)$^*$  & auto-cor  &  SDSS  &30,239  & $1.0-25.0$  &
       $5.45^{+0.35}_{-0.45}$      &      $1.90^{+0.04}_{-0.03}$    &       1.27       &
       $2.30^{+0.15}_{-0.20}$\\ [5pt] 
       Ivashchenko   et  al.   (2010)$^*$  &
       auto-cor  &  SDSS  &   52,303  &  $1.0-35$  &  $5.85\pm0.33$  &
       $1.87\pm0.07$ &  1.47 & $2.60\pm0.17$\\ [5pt]
       \hline
\label{table:studies}
\end{tabular}}
\end{table*}

\section{Comparison with Models of AGN Fueling and Evolution}

This paper  explores the environment of moderate  luminosity X-ray AGN
at $z\approx0.1$  by estimating their  cross-correlation function with
the SDSS Main Galaxy spectroscopic sample. The adopted methodology has
advantages  compared  to auto-correlation  function  studies, in  that
random  and systematic  uncertainties can  be better  controlled.  Our
clustering  methodology, AGN selection  function and  X-ray luminosity
range are very similar to those of Coil et al.  (2009) at $z\approx1$.
This allows direct comparison of the clustering properties of moderate
luminosity  AGN across redshift  by minimising  differential selection
effects. Our  results combined with those  of Coil et  al. (2009) show
that moderate luminosity X-ray selected  AGN live in dark matter halos
with masses $\rm M_{DMH}\approx  10^{13} \rm \, h^{-1} \,M_{\odot}$ at
all  redshifts since $z\approx1$.   If powerful  UV/optically selected
QSOs reside in lower mass halos  (few times $\rm 10^{12} \rm \, h^{-1}
\,M_{\odot}$), as suggested by some  studies (e.g. Croom et al.  2005,
Myers et al.  2007, da $\hat{A}$ngela  et al.  2008, Ross et al. 2009,
but see Figure \ref{fig:bias_dmh}),  then the fueling mode of
those sources may be different from that of the X-ray selected moderate 
luminosity AGN studied in this paper.

Semi-analytic models for the growth of SMBH that assume mergers as the
main mechanism for triggering AGN activity, predict parent dark matter
halos similar  to those determined for UV/optically  selected QSOs and
lower  than those  estimated here  for moderate  luminosity  X-ray AGN
(e.g.  Marulli  et al. 2008, Bonoli  et al.  2009).   Thus our results
argue against major  mergers as the main channel  for fueling the SMBH
in  moderate luminosity  AGN. Allevato et  al.  (2011)
have recently  expanded this conclusion to  higher X-ray luminosities,
$L_X \approx 10^{44}\,  \rm erg \, s^{-1}$.  They used
the  XMM-COSMOS field  to  estimate  dark matter  halos  in excess  of
$10^{13} \rm  \, h^{-1} \,M_{\odot}$ for powerful  X-ray sources. This
suggests that  a substantial fraction of the accretion density in the
Universe is associated with dark  matter  haloes with  masses
$\approx10^{13}  \rm  \,  h^{-1} 
\,M_{\odot}$, higher than what merger models predict.

Stochastic  accretion  as described  in  Hopkins  \& Hernquist  (2006)
cannot explain  the massive dark  matter halos of  moderate luminosity
AGN. In this  model, disk instabilities or minor  interactions fuel at
high accretion  rates relatively small  SMBHs in spiral  galaxies with
abundant cold  gas supply. This mechanism produces  by design moderate
luminosity AGN,  which however,  are predicted to  lie in  low density
environments,  as  this is  where  large  gas  reservoirs, i.e.   blue
galaxies, are typically found, particularly at low redshift.

A  more promising  alternative  is  the model  of  Ciotti \&  Ostriker
(2001), in which  stellar winds from evolved stars  in quiescent early
type  galaxies provide  the fuel  to  supply the  SMBH. The  accretion
energy heats the ambient gas,  slowing down subsequent infall when the
Compton temperature of  the emitted radiation is higher  than the mean
galactic gas  temperature. Evolution in this case  is characterized by
strong oscillations, in  which very fast and energetic  AGN bursts are
followed by  longer periods  during which the  SMBH is  dormant.  This
fueling mode applies to elliptical galaxies, which reside, on average,
in dark matter  halos similar to those of  our moderate luminosity AGN
sample.

One  approach for  getting  insights into  the  growth of  SMBH is  to
investigate how  the baryonic matter  evolves within the  typical dark
matter  haloes  that  AGN  are  found.  Empirical  methods  have  been
developed recently  to connect galaxies to haloes  at different epochs
to constrain the history of galaxy assembly (e.g.  Zheng et al.  2007,
Conroy \& Wechsler  2009, Zehavi et al.  2011,  Avila-Reese \& Firmani
2011).  Although these methods  cannot directly probe the formation of
SMBH at the  centres of galaxies, they can  provide useful information
on the conditions under which black holes grow.

A key  result from the  studies above is  that the ratio of  galaxy to
halo mass  shows a peak  at the dark  matter mass scale of  about $\rm
10^{12}  \, h^{-1}  \rm  \,M_{\odot}$, which  depends  only mildly  on
redshift, and  then decreases at higher/lower  masses (e.g.  Leauthaud
et  al.   2011,  Conroy  \&  Wechsler  2009).   The  halo  mass  where
star-formation  is most  efficient  is close to  that measured  for
luminous  UV/optically selected  QSOs, which  are proposed  to  be the
products  of gaseous  major mergers  (e.g.   Di Matteo  et al.   2005,
Hopkins et al.  2008, Marulli et  al.  2008, Bonoli et al.  2009).  In
contrast,  moderate luminosity  X-ray  AGN at  both $z\approx0.1$  and
$z\approx1$  (e.g. Coil  et al.   2009) live  in haloes  that  are, on
average,  offset from  the  sites where  star-formation proceeds  more
efficiently.   The typical  dark matter  halo of  these  systems ($\rm
10^{13}  \rm \, h^{-1}  \,M_{\odot}$) corresponds  to the  scale where
satellites start to dominate the stellar mass budget of the halo (e.g.
Leauthaud et  al.  2011),  i.e. small and  moderate size  groups (e.g.
Coil  et al.   2004).  This  might indicate  that  moderate luminosity
X-ray AGN may be fueled by  the cooling of hot gas associated with the
group environment (e.g.  Croton et al.  2006, Bower et al.  2006).
Semi-analytic models that include prescriptions for the growth of SMBHs
predict that this accretion mode can 
produce luminosities comparable to those of the  AGN sample presented
here (e.g.  Fanidakis et al.  2011).  Alternatively, the increasing
importance of satellites 
in  the  integrated  stellar  mass  of  $\rm  10^{13}  \rm  \,  h^{-1}
\,M_{\odot}$  haloes  suggests  that  they  may play  a  role  in  the
activation of the  SMBH, e.g.  via tidal interactions  or mergers with
the central  galaxy of the  halo.  Semi-analytic models
  of galaxy  evolution (e.g.   Bower et  al. 2006,  De  Lucia \&
Blaizot 2007) predict that the main route of stellar mass growth since
$z=1$ of  the central  galaxies of haloes  with present-day  mass $\rm
10^{13} \rm \, h^{-1} \,M_{\odot}$ are mergers with satellites (Zehavi
et al.  2011).  As long as  AGN are associated with the central galaxy
of  a  halo  (e.g.    Starikova  et  al.   2010),  such  gravitational
interactions  may  also be  responsible  for  triggering accretion  of
material onto the central SMBH.

We caution however, that studies  that connect stellar and halo masses
in a phenomenological way find contradictory results on the importance
of mergers in haloes with  present-day mass $\rm 10^{13} \rm \, h^{-1}
\,M_{\odot}$.  Conroy  \& Weschsler (2009)  argue that in  such haloes
star-formation is primarily responsible  for the growth of the stellar
mass of the central galaxies  since $z\approx1$. In contrast, Zheng et
al.  (2007) find that mergers with smaller galaxies, either satellites
or central galaxies of smaller haloes, drive the stellar mass assembly
since $z\approx1$ in those  haloes, in qualitative agreement with SAMs
(Zehavi et  al. 2011).  The discrepancy  is likely related  to the way
galaxies are  associated with dark matter haloes  in different studies
and  the  assumptions  adopted   to  determine  the  contributions  of
star-formation and merging to the stellar mass assembly. 

A limitation of schemes that connect galaxies to dark matter haloes is
that they  describe the average stellar  mass growth as  a function of
both galaxy and  halo mass.  There is however,  evidence that moderate
luminosity AGN hosts are different  from the typical galaxy that lives
in haloes similar  to those of AGN.  Using the  results of Behroozi et
al.  (2010, their Table 3) we find that dark matter haloes with masses
of  $\rm \approx  10^{13} \rm  \, h^{-1}  \,M_{\odot}$ at  $z=0.1$ and
$z=1$ are predicted to have  central galaxies with mean stellar masses
of  about $\rm  5\times 10^{10}  \rm \,  h^{-2} \,M_{\odot}$  and $\rm
6\times 10^{10} \rm \, h^{-2} \,M_{\odot}$, respectively.  The scatter
in stellar mass  at a fixed halo mass is  about 0.16\,dex (Behroozi et
al. 2010). Moderate  luminosity X-ray AGN hosts however,  have a broad
stellar mass distribution (e.g.  Shi et al.  2008, Bundy et al.  2008,
Georgakakis et al.   2011).  Using the stellar mass  function of X-ray
AGN from Georgakakis et al. (2011) for  example, we estimate mean
stellar  masses  of  about   $\rm  1.5\times  10^{10}  \rm  \,  h^{-2}
\,M_{\odot}$ at  $z\approx0.1$ and $\rm 2\times 10^{10}  \rm \, h^{-2}
\,M_{\odot}$ $z\approx0.8$  with a variance of about  0.4\,dex at both
redshifts.   This indicates  that  X-ray AGN,  for  their dark  matter
haloes,  live in  smaller than  average galaxies  in terms  of stellar
mass. X-ray  AGN  hosts have therefore experienced  less
merging  and/or less  star-formation (possibly due to feedback
processes associated with the central engine) compared  to typical
galaxies in 
dark matter haloes of similar mass. Alternatively, this  might be 
interpreted as evidence that a fraction of the X-ray 
AGN are not  associated with the central galaxies  of their haloes but
with satellites, which  are expected to have a  much wider stellar mass
function (e.g.   Leauthaud et  al.  2011).  Recent studies on
the Halo Occupation Distribution of AGN suggest that at least some of
them live in sattelite galaxies (e.g. Padmanabhan et ak. 2009, Miyaji
et al, 2011). Contrary to that, Starikova et al. (2010) find that X-ray
AGN reside close to the centres of their dark matter haloes and tend
to avoid satellites.

Evidence  that X-ray AGN  are in  lower mass  galaxies for  their dark
matter haloes  already exists in  the literature. Coil et  al.  (2009)
find that galaxies hosting X-ray AGN are more likely to reside in more
massive dark  matter halos than  the overall galaxy population  of the
same color and optical luminosity.  Digby  North et al. (2011) also
found that X-ray AGN live  on average in higher density  environments
compared to stellar-mass matched galaxy samples.

\section{Conclusions}

The XMM/SDSS  survey is used  to constrain the clustering  of moderate
luminosity ($L_X(\rm  2-10\,keV)=10^{41}-10^{44}\,erg \,s^{-1}$) X-ray
selected AGN at $z\approx0.1$. These  sources are found to  reside in
haloes  with  mass  of  about  $\rm  \approx  10^{13}  \rm  \,  h^{-1}
\,M_{\odot}$.   Comparison with  studies at  higher  redshift, suggest
that this  halo mass scale  corresponds to the typical  environment of
X-ray  AGN  in   the  luminosity  interval  above  at   least  out  to
$z\approx1$. 

Haloes  with masses $\rm  \approx 10^{13}  \rm \,  h^{-1} \,M_{\odot}$
correspond to the group  scale, where satellites dominated the stellar
mass budget  of the halo.  This  suggests that either  accretion of hot
gas  associated with  the  group environment  or  interactions of  the
central galaxy  with satellites are responsible for  the activation of
the SMBH in our AGN sample.

Splitting the AGN sample by  colours shows that those with red colours
are more  clustered than those in  blue colours. This  is in agreement
with  results at  higher  redshift,  $z\approx1$.  We  do  not find  a
dependence  of clustering  on obscuration  or accretion  luminosity. A
wider luminosity baseline than that  spanned by our sample is probably
needed to explore trends of the AGN clustering with $L_X$. 

There is also evidence that the hosts of moderate luminosity X-ray AGN
have lower stellar masses compared  to the typical galaxy in haloes of
the same size.  This  may  have important
implications for understanding the conditions under which supermassive
black holes at the centres of galaxies grow.

\section{Acknowledgments}
The  authors  acknowledge   financial  support  from  the  Marie-Curie
Reintegration  Grant  PERG03-GA-2008-230644.   Funding for  the  Sloan
Digital Sky  Survey (SDSS) has been  provided by the  Alfred P.  Sloan
Foundation, the  Participating Institutions, the  National Aeronautics
and Space  Administration, the  National Science Foundation,  the U.S.
Department of Energy, the  Japanese Monbukagakusho, and the Max Planck
Society.   The SDSS  Web site  is http://www.sdss.org/.   The  SDSS is
managed  by  the  Astrophysical  Research  Consortium  (ARC)  for  the
Participating  Institutions.  The  Participating Institutions  are The
University of Chicago, Fermilab, the Institute for Advanced Study, the
Japan Participation  Group, The  Johns Hopkins University,  Los Alamos
National  Laboratory, the  Max-Planck-Institute for  Astronomy (MPIA),
the  Max-Planck-Institute  for Astrophysics  (MPA),  New Mexico  State
University, University of Pittsburgh, Princeton University, the United
States Naval Observatory, and the University of Washington.

\vspace{10 mm}

\noindent{\bf References}

\vspace{5 mm}

\noindent
Abazajian K. N., et al., 2009, ApJS, 182, 543

\noindent
Aird J., et al., 2010, MNRAS, 401, 2531

\noindent
Akylas A., Georgantopoulos I., Plionis M., 2000, MNRAS, 318, 1036

\noindent
Allevato et al., 2011, ApJ accepted, preprint (astroph/1105.0520)

\noindent
Avila-Reese V., Firmani C., 2011, RevMexAA, (astroph/1103.4329)

\noindent
Barger A. J. et al., 2003, ApJ, 584, 61

\noindent
Basilakos S., Georgakakis A., Plionis M., Georgantopoulos I., 2004, ApJ, 607, L79 

\noindent
Basilakos S., Plionis M., Georgakakis A., Georgantopoulos I., 2005, MNRAS, 356, 183

\noindent
Behroozi P. S., Conroy C., Wechsler R. H., 2010, ApJ, 717, 379

\noindent
Blanton M. R., Roweis S., 2007, AJ, 133, 734

\noindent
Blanton M. R., 2006, ApJ, 648, 268

\noindent
Blanton M. R., et al., 2005, AJ, 129, 2562

\noindent
Bonoli S., et al., 2009, MNRAS, 396, 423

\noindent
Bower R., et al., 2006, ApJ, 648, 127

\noindent
Brusa M., et al., 2009, ApJ, 693, 8

\noindent
Bundy K., et al., 2008, ApJ, 681, 931

\noindent
Cannon R., et al., 2006, MNRAS, 372, 425

\noindent
Cappelluti N., et al., 2010, ApJ, 716, 209

\noindent
Ciotti L., Ostriker  J. P., 2007, ApJ, 665, 1038

\noindent
Coil A.L., et al., 2007, ApJ, 654, 115

\noindent
Coil A.L., et al., 2008, ApJ, 672, 153

\noindent
Coil A.L., et al., 2009, ApJ, 701, 1484

\noindent
Conroy C. \& Wechsler R. H., 2009, ApJ, 696, 620 

\noindent
Croom S. M., et al., 2004, MNRAS, 349, 1397

\noindent
Croom S. M., et al., 2005, MNRAS, 356, 415

\noindent
Croom S. M., et al., 2009, MNRAS, 392, 19

\noindent
Croton D. J., et al., 2006, MNRAS, 365, 11

\noindent
da $\hat{A}$ngela J., et al., 2008, MNRAS, 383, 565

\noindent
Degraf C., Di Matteo T., Springel V., 2010, MNRAS, 402, 1927

\noindent
De Lucia G. \& Blaizot J., 2007, MNRAS, 375, 2

\noindent
Di Matteo T., Springel V., Hernquist L., Nature, 433, 604

\noindent
Ebrero J., Mateos S., Stewart G. C., Carrera F. J., Watson M. G., 2009, A\&A, 500, 749

\noindent
Fanidakis, N., 2010, Ph.D. Thesis, Durham University

\noindent
Fanidakis N., et al., 2011, MNRAS, 410, 53

\noindent
Ferrarese L. \& Merritt D., 2000, ApJ, 539, 9

\noindent
Gebhardt K. et al., 2000, ApJ, 543, 5

\noindent
Georgakakis A., et al., 2008, MNRAS, 391, 183

\noindent
Georgakakis A., et al, 2009, MNRAS, 397, 623

\noindent
Georgakakis A. \& Nandra K., 2011, MNRAS, 414, 992

\noindent
Georgakakis A. et al., 2011, MNRAS in press, arxiv1109.0287

\noindent
Gilli R., et al. 2005, A\&A, 430, 811 

\noindent
Gilli R., et al. 2009, A\&A, 494, 33

\noindent
Hickox R.C., et al., 2009, ApJ, 696, 891

\noindent
Hickox R.C., et al., 2011, ApJ, 731, 117

\noindent
Hopkins P. F., Hernquist, L., 2006, ApJS, 166, 1

\noindent
Hopkins P. F., et al., 2007, ApJ, 662, 110

\noindent
Hopkins P. F., Hernquist L., Cox T.  J., Kere D., 2008, ApJS, 175, 356

\noindent
Ivashchenko G., Zhdanov V. I., Tugay, A. V., 2010, MNRAS, 409, 1691

\noindent
Kenter A., 2005, ApJS, 161, 9

\noindent 
King A., 2005, ApJ, 635, 121

\noindent
Kormendy J. \& Richstone D., 1995, ARA\&A, 33, 581

\noindent
Krumpe M., Miyaji T., Coil A. L., 2010, ApJ, 713, 558

\noindent
Leauthaud A., et al., 2011, ApJ accepted, preprint (astroph/1103.2077)

\noindent
Leauthaud A., et al., 2011, ApJ submitted, preprint (astroph/1104.0928)

\noindent
Limber D. N., 1953, ApJ, 117, 134

\noindent
Madgwick D. S., et al., 2003, MNRAS, 344, 847

\noindent
Marulli F., Crociani D., Volonteri M., Branchini E., Moscardini L., 2006, MNRAS, 368, 1269

\noindent
Marulli F., Bonoli S., Branchini E., Moscardini L., Springel V., 2008, MNRAS, 385, 1846

\noindent
Marulli F., et al., 2009, MNRAS, 396, 1404

\noindent
Miyaji T., et al., 2007, ApJS, 172, 396

\noindent
Miyaji T., Krumpe M., Coil A. L., Aceves H., 2011, ApJ, 726, 83

\noindent
Mo H. J., White S. D. M., 1996, MNRAS, 282, 347

\noindent
Mountrichas G., Sawangwit U., Shanks T., Croom S. M., Schneider D. P., Myers A. D., Pimbblet K., 2009, MNRAS, 394, 2050

\noindent
Mullis C. R., et al., 2004, ApJ, 617, 192

\noindent
Myers A. D., et al., 2005, MNRAS, 359, 741

\noindent
Myers A. D., et al., 2007, ApJ, 658, 85

\noindent
Myers, A. D., et al. 2007, ApJ, 658, 99

\noindent
Nandra K., Pounds K. A., 1994, MNRAS, 268, 405

\noindent
Padmanabhan, N., et al., 2008, ApJ, 674, 1217

\noindent
Padmanabhan, N., et al., 2009, MNRAS, 397, 1862

\noindent
Peebles P. J. E., 1979, MNRAS, 189, 89

\noindent
Peebles P. J. E., 1980, The Large-scale Structure of the Universe. Research
supported by the National Science Foundation. Princeton Univ. Press,
Princeton, NJ, p. 435

\noindent
Plionis M., Rovilos M., Basilakos S., Georgantopoulos I., Bauer F., 2008, ApJ, 674, 5

\noindent
Richards G. T., et al., 2002, AJ, 123, 2945

\noindent
Ross N. P., et al., 2007, MNRAS, 381, 573

\noindent
Ross N. P., et al., 2008, MNRAS, 387, 1323

\noindent
Ross N. P., et al., 2009, ApJ, 697, 1634

\noindent
Saunders W., Rowan-Robinson M., Lawrence A., 1992, MNRAS, 258, 134

\noindent
Schneider D. P., 2005, AJ, 130, 367

\noindent
Schlegel D. J., Finkbeiner D. P., Davis M., 1998, ApJ, 500, 525

\noindent
Shankar, F., Crocce M., Miralda-Escud\'{e} J., Fosalba P., Weinberg D.H., 2010, ApJ, 718, 231

\noindent
Shanks T., Croom S. M., Fine S., Ross N. P., Sawangwit U., 2011, MNRAS submitted, preprint (astroph/1105.2547v1)

\noindent
Sheth R. K., Mo H. J., Tormen G., 2001, MNRAS, 323, 1

\noindent
Shi Y., et al., 2008, ApJ, 688, 794

\noindent
Silk J. \& Rees M. J., 1998, A\&A, 331, 1 

\noindent
Smith R. E., et al., 2003, MNRAS, 341, 1311

\noindent
Spergel D. N., et al., 2007, ApJS, 170, 377

\noindent
Springel V., et al., 2005, Nat, 435, 629

\noindent
Starikova S., et al., 2010, ApJ submitted, preprint (astroph/1010.1577)

\noindent
Strauss M. A., et al., 2002, AJ, 124, 1810

\noindent
Tueller J., et al., 2010, ApJS, 186, 378

\noindent
van den Bosch, F.C. 2002, MNRAS, 331, 98

\noindent
Yang Y., Mushotzky R. F., Barger A. J., Cowie L. L., 2006, ApJ, 645, 68

\noindent
Zehavi I., et al., 2005, ApJ, 630, 1

\noindent
Zehavi I., Patiri S., Zheng, Z., 2011, ApJ submtted, preprint (astroph/1104.0389)

\noindent
Zheng Z., Coil A. L., Zehavi I., 2007, ApJ, 667, 760

\vspace{6 mm}

\end{document}